\documentclass[aps,journal=prb,twocolumn,noeprint]{revtex4-1}
\usepackage{graphicx}
\usepackage{epstopdf}
\usepackage{bmpsize}
\usepackage{amsmath,amsfonts,amsthm,bm}
\usepackage{bigints}
\usepackage{tikz}
\usetikzlibrary{decorations.pathreplacing}
\usepackage{subcaption}
\usepackage[bookmarks=false]{hyperref}
\hypersetup{colorlinks=true, citecolor=blue, urlcolor=blue, linkcolor=blue}

\makeatletter
\renewcommand{\maketag@@@}[1]{\hbox{\m@th\normalsize\normalfont#1}}%
\makeatother

\def\!{\mskip-\thinmuskip}

\begin{document}

\title{Current-induced forces in nanosystems: A hierarchical equations of motion approach}
\author{Samuel L. Rudge}
\affiliation{Institute of Physics, University of Freiburg, Hermann-Herder-Strasse 3, 79104 Freiburg, Germany}
\author{Yaling Ke}
\affiliation{Institute of Physics, University of Freiburg, Hermann-Herder-Strasse 3, 79104 Freiburg, Germany}
\author{Michael Thoss}
\affiliation{Institute of Physics, University of Freiburg, Hermann-Herder-Strasse 3, 79104 Freiburg, Germany}

\begin{abstract}
\noindent A new approach to calculating current-induced forces in charge transport through nanosystems is introduced. Starting from the fully quantum mechanical hierarchical equations of motion formalism, a timescale separation between electronic and vibrational degrees of freedom is used to derive a classical Langevin equation of motion for the vibrational dynamics as influenced by current-induced forces, such as the electronic friction. The resulting form of the friction is shown to be equivalent to previously derived expressions. The numerical exactness of the hierarchical equations of motion approach, however, allows the investigation of transport scenarios with strong intrasystem and system-environment interactions. As a demonstration, the electronic friction of three example systems is calculated and analyzed: a single electronic level coupled to one classical vibrational mode, two electronic levels coupled to one classical vibrational mode, and a single electronic level coupled to both a classical and quantum vibrational mode.
\end{abstract}

\maketitle

\section{Introduction}

A comprehensive picture of nonequilibrium charge transport through nanosystems, such as molecular junctions \cite{Tal2008,Secker2011}, nanoelectromechanical systems \cite{Craighead2000,Ekinci2005}, or suspended carbon nanotubes \cite{Weig2004,Sapmaz2006}, requires not only a description of charge effects but also the coupling of electronic and vibrational degrees of freedom. Novel effects arising from such electron-vibration interactions include the Franck-Condon blockade \cite{Koch2005,Koch2006a,Koch2006b,Leturcq2009}, current-induced heating and cooling \cite{Haertle2011a,Haertle2018,Haertle2011b,Kaspar2022}, and non-renewal statistics \cite{Koch2005,Kosov2017b,Rudge2019a,Schinabeck2014}. Including such interactions, however, requires transport methods able to treat the quantum nature of the nanosystem while simultaneously incorporating nonequilibrium effects from the electrodes. Examples of such methods include nonequilibrium Green's functions (NEGFs) \cite{Mitra2004,Frederiksen2007,Galperin2007,Haertle2008,Novotny2011,Dash2011,Agarwalla2015,Ryndyk2006,Martin-Rodero2008,Entin-Wohlman2009,Egger2008,Galperin2004,Tikhodeev2001,Paulsson2005}, which can treat strong molecule-lead couplings, and Born-Markov master equations (BMMEs) \cite{Koch2006a,Li2005b,Leijnse2010,Rudge2021}, which can treat strong intra-system interactions. In recent years, furthermore, many more sophisticated and numerically exact methods have also been developed, including path integral approaches \cite{Muehlbacher2008,Weiss2008}, the multilayer multiconfiguration time-dependent Hartree (ML-MCTDH) approach \cite{Wang2011,Wang2016} or the hierarchical equations of motion (HEOM) approach \cite{Schinabeck2016,Schinabeck2018,Schinabeck2020,Jin2008}, although these can be computationally challenging. 

In the specific context of transport through molecular junctions, many vibrational degrees of freedom, each corresponding to the motion of atomic nuclei, are necessary for realistic simulations. In these scenarios, mixed quantum-classical approaches, where the vibrational dynamics are treated classically while subject to current-induced forces from the quantum electronic degrees of freedom, are often more viable \cite{Tully1990,Stock1997,Martinez1996}. When the nuclear motion is slow or limited to small displacements, furthermore, the vibrational degrees of freedom follow a Langevin equation of motion. In such regimes,  the current-induced force splits into a zeroth-order term, corresponding to the Born-Oppenheimer approximation, and a first-order correction linear in the vibrational momenta or coordinates \cite{Dou2018a}: the electronic friction, which is the specific current-induced force of interest in this study. Because the current-induced forces offer insight into the role of the electron-vibration interaction on the dynamics \cite{Dzhioev2011,Dzhioev2013,Preston2021,Mozyrsky2006,Kershaw2020,Bode2012,Lue2010,Lue2012,Todorov2011,Todorov2014,Maurer2016,Askerka2016,Bustos-Marun2013,Fernandez-Alcazar2019,Calvo2017}, especially in the regime of vibrational instabilities \cite{Sabater2015,Ring2020}, various theories have been developed for calculating them in the context of quantum transport through nanostructures. 

These approaches can be broadly categorized into two classes: those that derive the electronic friction under a weak electron-vibration coupling, and those that derive it using a timescale separation between electronic and vibrational degrees of freedom. To the first class belongs an influence functional approach that has been derived by several groups \cite{Hussein2010,Metelmann2011,Lue2012,Lue2010,Lue2011,Brandbyge1995,Caldeira1983,Schmid1982}. The second class contains a wider variety of transport methods. These include the approach introduced in the seminal paper by Head-Gordon and Tully \cite{Head-Gordon1995,Stock1997,Meyer1979}, which has since been extended to finite temperatures \cite{Dou2017c,Maurer2016} and has shown to be equivalent to a combined scattering theory and NEGF approach \cite{Bode2012,Dou2017a}. The connections between these various approaches and assumptions have been explicitly explored by Chen \textit{et al.} \cite{Chen2019a}, who also used their unified formalism to investigate interacting systems \cite{Chen2019b}. Although these approaches explicitly treat the vibrational degrees of freedom as classical, there have also been recent attempts to derive a quantum equivalent to the classical electronic friction \cite{Martinazzo2022a,Martinazzo2022b}.

Under the timescale separation assumption, one can also formulate master equation approaches, such as that proposed by Dou and Subotnik in Refs. \cite{Dou2015d,Dou2016a}. Here, a quantum-classical Liouville equation (QCLE) for the vibrational dynamics was embedded into a BMME for the electronic dynamics. This approach has the benefit of allowing for strong intra-system interactions, but is limited to weak molecule-lead couplings under the Born-Markov assumption. This method is particularly relevant to the purpose of this paper, in which a similar approach to calculating the electronic friction is introduced, although with the numerically exact HEOM approach instead of a BMME. Starting from the full HEOM method, in which the vibrational degrees of freedom are originally treated quantum mechanically, the QCLE is formed by applying a partial Wigner transform of the vibrational degrees of freedom \cite{Kapral1999,Imre1967}. Similar to Ref. \cite{Dou2016a}, the QCLE is then transformed to a Fokker-Planck equation in the adiabatic limit of slow vibrational motion, from which the Markovian electronic friction of the corresponding Langevin equation can be identified. The resulting electronic friction is then shown to be equivalent to expressions derived by treating all electronic degrees of freedom in the system and bath equally \cite{Dou2018a,Chen2019a}. Because it is derived from HEOM,  which systematically incorporates higher-order tunneling and non-Markovian effects for nonequilibrium charge transport, the results are general in- and out-of-equilibrium for systems with strong interactions and strong molecule-lead couplings.

In order to demonstrate the approach, three model systems are considered. In the first system, a single electronic level is linearly coupled to a classical vibrational degree of freedom. The electronic friction of this non-interacting system has previously been calculated from NEGFs \cite{Dou2018a}, which the HEOM approach reproduces. In the second system, electron-electron interactions are incorporated via an Anderson impurity, and the friction at equilibrium is shown to be equivalent to the combined numerical renormalization group (NRG) and NEGF method from Ref. \cite{Dou2017c}. The HEOM approach, however, can also investigate this model out-of-equilibrium, and the corresponding friction at finite bias voltages is analyzed. The final case study couples an electronic level to both a high-frequency mode, treated quantum mechanically in the HEOM framework, and a low-frequency mode, treated classically in the Langevin equation framework. This system is similar to one treated recently in Ref. \cite{Chen2019b}, where the high-frequency mode was used to model a strong light-matter interaction. The additional vibrational degree of freedom adds extra structure to the friction on the classical mode in nonequilibrium scenarios, even turning it negative for particular parameters. 

The paper is structured as follows. In Section \ref{sec: Model}, the general model of a molecular junction is outlined. Next, in Section \ref{sec: Quantum transport theory}, the two methods used to describe nonequilibrium charge transport are introduced, the HEOM and BMME approaches. In Section \ref{sec: Semi-classical transport theory}, these are then Wigner-transformed to a QCLE and, under the appropriate assumptions, to a Markovian Fokker-Planck equation, from which the friction can be identified. In order to demonstrate the method, Section \ref{sec: Results} then presents results for the three model systems, with the conclusions contained in Section \ref{sec: Conclusion}. Relevant derivations can be found in Appendices \ref{app: On the derivation of the Markovian Fokker-Planck equation}, \ref{app: Equivalence of the two friction tensors}, and  \ref{app: Non-Markovian friction tensor}.

Throughout the paper we use units where $\hbar = e = k_{B} = 1$.

\section{Model}\label{sec: Model}

The total Hamiltonian of a junction containing a nanosystem, $H$, can be decomposed into a part describing the electrons in the leads, $H_{\text{B}}$, a part containing the nanosystem, $H_{\text{S}}$, and a part containing the interaction between them, $H_{\text{SB}}$:
\begin{align}
H & = H_{\text{S}} + H_{\text{B}} + H_{\text{SB}}.
\end{align}

The system Hamiltonian contains several coupled electronic and vibrational degrees of freedom, which can generally be written as 
\begin{align}
H_{\text{S}} & = \sum_{mm'}\varepsilon_{mm'}(\hat{\bold{x}}) d_{m}^{\dag}d^{}_{m'} + \sum_{m,m' < m} U_{mm'}(\hat{\bold{x}})d_{m}^{\dag}d^{}_{m}d_{m'}^{\dag}d^{}_{m'} \nonumber \\
& \:\:\:\:\: + \sum_{i = 1}^{N}\frac{\hat{p}^{2}_{i}}{2m_{i}} + V_{\text{vib}}(\hat{\bold{x}}), \label{eq: general system Hamiltonian}
\end{align}
where $\left(\hat{\bold{x}} = (\hat{x}_{1},\dots,\hat{x}_{N}),\hat{\bold{p}} = (\hat{p}_{1},\dots,\hat{x}_{N})\right)$ are vectors of displacement operators and their corresponding conjugate momenta for each of the $N$ vibrational coordinates. The potential of the unperturbed vibrational degrees of freedom is $V_{\text{vib}}(\hat{\bold{x}})$. The vibrational motion is coupled to the electronic degrees of freedom via $\varepsilon_{mm'}(\hat{\bold{x}})$, which is the energy of the $m$th electronic energy level if $m = m'$, or the direct hopping between the $m$th and $m'$th electronic energy levels if $m \neq m'$, and $U_{mm'}(\hat{\bold{x}})$, which is the electron-electron interaction between electrons in the $m$th and $m'$th electronic energy levels. The annihilation and creation operators corresponding to the $m$th electronic energy level are $d^{}_{m}$ and $d^{\dag}_{m}$, respectively. Note that the vibrational operators have been explicitly written with operator notation to distinguish $\hat{\bold{x}}$ and $\hat{\bold{p}}$ from their classical variable counterparts, $\bold{x}$ and $\bold{p}$, which will be introduced in Sec. \ref{subsec: Quantum-classical Liouville equation}. 

Although the interest in current-induced forces and mixed quantum-classical dynamics is often from the perspective of transport through molecular junctions, the Hamiltonian in Eq.\eqref{eq: general system Hamiltonian} and the theory that follows in Secs. \ref{sec: Quantum transport theory} and \ref{sec: Semi-classical transport theory} is general for transport through any generic nanosystem.

The leads are modeled as two reservoirs of non-interacting electrons, labeled the left, $L$, and right, $R$, electrodes, 
\begin{align}
H_{\text{B}} & = \sum_{\alpha \in \{L,R\}}\sum_{k_{\alpha}} \varepsilon_{k_{\alpha}} c^{\dag}_{k_{\alpha}}c^{}_{k_{\alpha}}.
\end{align}
The $(k_{\alpha})$th level is formed through the corresponding annihilation and creation operators, $c^{}_{k_{\alpha}}$ and $c^{\dag}_{k_{\alpha}}$, and has energy $\varepsilon_{k_{\alpha}}$. Each electrode is held at local equilibrium, with temperatures $T_{\alpha}$ and chemical potentials $\mu_{\alpha}$. By introducing a temperature difference, $\Delta T = T_{L} - T_{R}$, or voltage bias, $\Phi = \mu_{L} - \mu_{R}$, across the junction, the system can be driven out of equilibrium, producing a non-zero total current as electrons transport across the junction. 

The interaction term in the total Hamiltonian, which is linear in both the system and bath annihilation and creation operators, describes electron transfer between the nanosystem and the electrodes,
\begin{align}
H_{\text{SB}} & = \sum_{\alpha,k_{\alpha}}\sum_{m} V_{k_{\alpha},m}\left(c^{\dag}_{k_{\alpha}}d^{}_{m} + d^{\dag}_{m}c^{}_{k_{\alpha}}\right),
\end{align}
where the molecule-lead couplings, $V_{k_{\alpha},m}$, are assumed to be real and independent of the vibrational positions. Although this assumption is standard in the literature and greatly simplifies the quantum and semi-classical transport theory, a more general approach would include $V_{k_{\alpha},m}^{}(\hat{\mathbf{x}})$, such as in Ref. \cite{Dou2017d}, which will be incorporated in future work. Finally, the spectral density of bath $\alpha$ is introduced,
\begin{align}
\Gamma_{\alpha,mm'}(\varepsilon) & = 2\pi \sum_{k_{\alpha}} V^{}_{k_{\alpha},m}V^{}_{k_{\alpha},m'}\delta(\varepsilon-\varepsilon_{k_{\alpha}}),
\end{align}
which describes the interaction strength between electronic levels in the nanosystem and electrodes. 

\section{Quantum transport theory}\label{sec: Quantum transport theory}

This section outlines the two transport methods used to simulate the nonequilibrium quantum dynamics of the general Hamiltonian in Eq.\eqref{eq: general system Hamiltonian}. Although both methods have been used extensively, and there exist several detailed reviews, brief outlines are included here for self-consistency within the paper. 

\subsection{Hierarchical equations of motion}\label{subsubsec: HEOM}

The HEOM formalism is a powerful method able to simulate nonequilibrium quantum transport with strong interactions and beyond the Born-Markov regime. Although the general steps are outlined in this section, readers interested in a comprehensive review are referred to the original work by Tanimura and Kubo \cite{Tanimura1989}, its extension to fermionic systems \cite{Tanimura2006,Haertle2015,Jin2007,Jin2008,Xiao2009,Yan2014,Wenderoth2016}, and the review articles in Refs. \cite{Tanimura2020,Ye2016}.

As with all master equation methods, the central object of interest in the HEOM approach is the reduced density matrix of the nanosystem, $\rho(t)$, which is obtained by tracing out the bath degrees of freedom from the total density matrix, $\rho(t) = \text{Tr}_{\text{B}}\left[\rho_{\text{T}}(t)\right]$. One could obtain $\rho(t)$ by solving the corresponding Liouville-von Neumann equation,
\begin{align}
\frac{\partial}{\partial t}\rho(t) & = -i\text{Tr}_{\text{B}}\left[H,\rho_{\text{T}}(t)\right].
\end{align}
This, however, is a difficult task; the dynamics involves not only the coherent time-evolution of the nanosystem, but non-trivial dissipation due to interaction with the electrodes. By employing a coherent state representation for the fermionic annihilation and creation operators, the HEOM approach treats the system-bath interaction via the Feynman-Vernon influence functional, which collects the bath effects into an effective action term: the influence functional. In this work, the influence functional is derived under the assumption of a factorized initial state, 
\begin{align}
\rho_{\text{T}}(0) & = \rho(0)\otimes\rho_{\text{B}}(0),
\end{align}
as initial correlations between the nanosystem and the bath require a more complicated imaginary-time propagation approach \cite{Tanimura2014,Ke2022}. 

As long as the molecule-lead coupling is linear in the bath annihilation and creation operators, one can then show that a cumulant expansion of the influence functional terminates exactly after the second term. The bath statistics are therefore Gaussian and their effect is exactly described with the two-time bath correlation functions,
\begin{align}
C^{\sigma}_{\alpha,mm'}(t - \tau) & = \sum_{k_{\alpha}} V^{}_{k_{\alpha},m}V^{}_{k_{\alpha},m'}\text{Tr}_{\text{B}}\left[c^{\sigma}_{k_{\alpha}}(t)c^{\bar{\sigma}}_{k_{\alpha}}(\tau)\rho_{\text{B}}(0)\right],
\end{align}
where the notation $\sigma \in \{+,-\}$, with $c^{-(+)}_{k_{\alpha}} = c^{(\dag)}_{k_{\alpha}}$, has been introduced. Based on the spectral densities of the baths, the bath-correlation functions can be rewritten as 
\begin{align}
C^{\sigma}_{\alpha,mm'}(t) & = \frac{1}{2\pi} \int^{\infty}_{-\infty}  d\omega \: e^{\sigma i \omega t} \Gamma_{\alpha,mm'}(\omega) f^{\sigma}_{\alpha}(\omega), \label{eq: spectral bath correlation}
\end{align}
with the Fermi-Dirac functions,
\begin{align}
f^{\sigma}_{\alpha}(\omega) & = \frac{1}{1 + e^{\sigma\beta(w - \mu_{\alpha})}},
\end{align}
and $\beta = 1/k_{B}T$. 

The next key step in deriving the HEOM is to assume that the bath-correlation functions can be expanded in a series of exponential functions,
\begin{align}
C^{\sigma}_{\alpha,mm'}(t) & = V^{}_{\alpha,m}V^{}_{\alpha,m'} \sum_{\ell = 0}^{\ell_{\max}} \eta_{\alpha,\sigma,\ell,m} e^{-\kappa_{\alpha,\sigma,\ell,m}t}. \label{eq: exponential decomposition}
\end{align}
By incorporating a sum-over-poles decomposition scheme for the Fermi-Dirac functions and choosing a specific form of the bath spectral densities, the coefficients, $\eta_{\alpha,\sigma,\ell,m}$, and exponents, $\kappa_{\alpha,\sigma,\ell,m}$, can then be calculated directly with residue theory from Eq.\eqref{eq: spectral bath correlation}. In this paper, the Pad\'{e} decomposition is exclusively used, which performs much bettter than a standard Matsubara decomposition \cite{Ozaki2007}.  All results below are also calculated under the wide-band limit, such that $\Gamma_{\alpha,mm'}(\varepsilon) = \Gamma_{\alpha,mm'} = 2\pi V^{}_{\alpha,m}V^{}_{\alpha,m'}$, where the constant density of states has been absorbed into the definition of $V_{\alpha,m}$. Although the wideband limit is a physically reasonable assumption for many systems of interest, such as molecules attached to metal electrodes \cite{Verzijl2013}, it is not actually necessary for the HEOM approach. The decomposition of the bath-correlation functions in Eq.\eqref{eq: spectral bath correlation} can be performed for a wide variety of $\Gamma_{\alpha,mm'}(\varepsilon)$, particularly with the recent development of numerical schemes able to treat density of state functions without an explicit analytic decomposition \cite{Xu2022,Dan2022}.

By repeatedly differentiating the bath-correlation functions, a hierarchy of first-order differential equations coupling the reduced system density matrix to a series of auxiliary density operators (ADOs), $\rho^{(n)}_{\mathbf{j}}$, is formed,
\begin{align}
\frac{\partial}{\partial t}\rho^{(n)}_{\mathbf{j}}(t) & = -i\left[H_{\text{S}},\rho^{(n)}_{\mathbf{j}}\right]  - \left(\sum_{r = 1}^{N} \kappa^{}_{j_{r}}\right)\rho^{(n)}_{\mathbf{j}} \nonumber \\
&  - i\sum_{r=1}^{n} (-1)^{n-r}\mathcal{C}_{j_{r}}\rho^{(n-1)}_{\mathbf{j}^{-}} - i\sum_{j} \mathcal{A}^{\bar{\sigma}}_{\alpha,m} \rho^{(n+1)}_{\mathbf{j}^{+}}. \label{eq: HEOM Final}
\end{align}
The notation here uses a vector of condensed super-indices, $\mathbf{j} = \left(j_{n} \dots j_{1}\right)$ with $j_{r} = \{\alpha_{j_{r}},\sigma_{j_{r}},\ell_{j_{r}},m_{j_{r}}\}$. The ADO tier, which is given the label $n$, is then the number of indices required to form the $\rho^{(n)}_{\mathbf{j}}$ ADOs. In this formulation, the $0$th-tier ADO is then just the reduced density matrix of the nanosystem, $\rho^{(0)}(t) = \rho(t)$. The HEOM also contain two more super-indices, $\mathbf{j}^{-}  = \left(j_{n},\dots , j_{r+1},j_{r-1},\dots, j_{1}\right)$, which is formed by removing an ADO, and $\mathbf{j}^{+} = \left(j,j_{n},\dots, j_{1}\right)$, which is formed by adding one. The superoperators $\mathcal{C}_{j_{r}}$ and $\mathcal{A}^{\bar{\sigma}}_{\alpha,m}$, meanwhile, are defined by their action on the ADOs,
\begin{align}
\mathcal{C}_{j_{r}}\rho^{(n)}_{\mathbf{j}}(t) & = V^{}_{\alpha,m}\left(\eta^{}_{j_{r}}d^{\sigma}_{m}\rho^{(n)}_{\mathbf{j}}(t) - (-1)^{n}\eta^{*}_{j_{r}}\rho^{(n)}_{\mathbf{j}}d^{\sigma}_{m}(t)\right) \label{eq: coupling down superoperator} \\
\mathcal{A}^{\bar{\sigma}}_{\alpha,m}\rho^{(n)}_{\mathbf{j}}(t) & = V^{}_{\alpha,m} \left(d^{\bar{\sigma}}_{m}\rho^{(n)}_{\mathbf{j}}(t) + (-1)^{n}\rho^{(n)}_{\mathbf{j}}d^{\bar{\sigma}}_{m}(t)\right). \label{eq: coupling up superoperator}
\end{align}

Eq.\eqref{eq: HEOM Final} represents a large set of coupled $1$st-order differential equations. In order to generate the hierarchy, one first constructs the equation of motion for the reduced density matrix of the nanosystem, which couples according to Eq.\eqref{eq: coupling up superoperator} to the $1$st-tier ADOs, $\rho^{(1)}_{j}(t)$. The respective $1$st-tier equations of motion are then generated from Eq.\eqref{eq: HEOM Final}, which couple according to Eqs.\eqref{eq: coupling down superoperator} and \eqref{eq: coupling up superoperator} to the reduced density matrix and $2$nd-tier ADOs, $\rho^{(2)}_{j_{2}j_{1}}$, and reduced density matrix, respectively. This process continues until the hierarchy naturally terminates at some, typically large, tier. In practice, the hierarchy must be truncated before this point due to numerical constraints, but at a large enough tier such that the dynamics are converged. In this work, a simple scheme is used that truncates the hierarchy directly at a maximum tier, $N$, such that the $\{\rho^{(N)}_{\mathbf{j}}(t)\}$ couple only to ADOs of the same tier and ADOs of the tier below, $\{\rho^{(N-1)}_{\mathbf{j}^{-}}(t)\}$. 
The resulting coupled set of $1$st-order differential equations, with the appropriate initial condition, is solved with a $4$th-order Runge-Kutta method.

\subsection{Born-Markov master equation}\label{subsubsec: Born-Markov rate equation}

In order to compare to the method in Ref. \cite{Dou2016a}, a Born-Markov master equation is also used to calculate the electronic friction. Unlike the numerically exact HEOM method, Born-Markov theory expands the Liouville-von Neumann equation to second-order in the system-bath interaction and explicitly assumes that bath-correlations decay on a timescale quicker than the dynamics of the nanosystem \cite{Li2005b,Rudge2021}. The resulting equation of motion for the reduced system density matrix,
\begin{align}
\frac{\partial}{\partial t}\rho(t) & = -i\left[H_{\text{S}},\rho(t)\right] \nonumber \\
& \:\:\:\:\:\: - \int^{\infty}_{0} d\tau \: \text{Tr}_{\text{B}}\left[H_{\text{SB}},\left[H_{\text{SB}}(\tau),\rho(t)\otimes\rho^{\text{eq}}_{\text{B}}\right]\right], \label{eq: BMME final}
\end{align}
is closed and, therefore, easier to treat numerically, but it is only valid in the weak molecule-lead coupling regime. In Eq.\eqref{eq: BMME final}, the system-bath interaction follows the time-evolution $H_{\text{SB}}(\tau) = e^{-i(H_{\text{S}}+H_{\text{B}})\tau}H_{\text{SB}}e^{i(H_{\text{S}}+H_{\text{B}})\tau}$. In all calculations, the BMME will be further simplified by excluding the off-diagonal elements of the density matrix from the transport, yielding essentially a rate equation between pure states of the nanosystem \cite{Rudge2019b}. 

\section{Quantum-classical transport theory} \label{sec: Semi-classical transport theory}

\subsection{Quantum-classical Liouville formulation of the HEOM}\label{subsec: Quantum-classical Liouville equation}

In this subsection, the fully quantum HEOM approach is transformed into a quantum-classical Liouville equation. First, the reduced system density matrix is partially Wigner transformed with respect to the vibrational degrees of freedom \cite{Kapral1999,Dou2016a},
\begin{align}
\rho^{\text{el}}(\mathbf{x},\mathbf{p} ; t) & = (2\pi)^{-3N} \int d\mathbf{y} \: e^{i\mathbf{p}\cdot\mathbf{y}}\langle \mathbf{x} - \frac{\mathbf{y}}{2}| \rho(t) | \mathbf{x} + \frac{\mathbf{y}}{2} \rangle.
\end{align}
Note that the Wigner transformed quantities, such as $H^{\text{el}}_{S}(\mathbf{x},\mathbf{p})$, are evaluated at positions and momenta, $(\mathbf{x},\mathbf{p})$, but are still operators in the electronic subspace of the nanosystem.

Partially Wigner transforming the first of the coupled differential equations in the HEOM yields
\begin{align}
\frac{\partial}{\partial t}\rho^{\text{el}}(\mathbf{x},\mathbf{p},t) & = -i\left(\left[H^{\text{el}}_{\text{S}},\rho(t)\right]\right)^{\text{el}}(\hat{\mathbf{x}},\hat{\mathbf{p}}) \nonumber \\
& \:\:\:\:\: - i\sum_{j} \mathcal{A}^{\bar{\sigma}}_{K} \rho^{(1),\text{el}}_{j}(\mathbf{x},\mathbf{p} ; t), \label{eq: zeroth tier HEOM Wigner transform}
\end{align}
where the partial Wigner transform of an $n$th-tier ADO is 
\begin{align}
\rho^{(n),\text{el}}_{\mathbf{j}}(\mathbf{x},\mathbf{p} ; t) & = (2\pi)^{-3N} \int d\mathbf{y} \: e^{i\mathbf{p}\cdot\mathbf{y}}\langle \mathbf{x} - \frac{\mathbf{y}}{2}| \rho^{(n)}_{\mathbf{j}}(t) | \mathbf{x} + \frac{\mathbf{y}}{2} \rangle.
\end{align}
The crucial step that transforms this into a quantum-classical equation comes when evaluating the partial Wigner transform of a product of operators \cite{Imre1967}, which is
\begin{align}
(AB)^{\text{el}}(\mathbf{x},\mathbf{p}) & = A^{\text{el}}(\mathbf{x},\mathbf{p})e^{\hbar\Lambda/2i}B^{\text{el}}(\mathbf{x},\mathbf{p}), \label{eq: Wigner transform product expansion}
\end{align}
with the Poisson bracket operator 
\begin{align}
\Lambda & = \overleftarrow{\nabla}_{\mathbf{p}}\cdot\overrightarrow{\nabla}_{\mathbf{x}} - \overleftarrow{\nabla}_{\mathbf{x}}\cdot\overrightarrow{\nabla}_{\mathbf{p}}.
\end{align}
The direction in which each operator acts is indicated by the arrows. In expanding the exponential, a classical approximation for the vibrational degrees of freedom is obtained by keeping only terms up to  linear in $\hbar$, such that the commutator in Eq.\eqref{eq: zeroth tier HEOM Wigner transform} becomes
\begin{align}
& \left(\left[H^{\text{el}}_{\text{S}},\rho(t)\right]\right)^{\text{el}}(\hat{\mathbf{x}},\hat{\mathbf{p}}) \approx \nonumber \\ & \hspace{0.5cm}\left[H^{\text{el}}_{\text{S}}(\mathbf{x},\mathbf{p}),\rho^{\text{el}}(\mathbf{x},\mathbf{p} ; t)\right] + i\left\{H^{\text{el}}_{\text{S}}(\mathbf{x},\mathbf{p}),\rho^{\text{el}}(\mathbf{x},\mathbf{p} ; t) \right\}_{a}, \label{eq: Wigner transform commutator approximation}
\end{align}
where $\{A_{1},A_{2}\}_{a} = \frac{1}{2}\left(\{A_{1},A_{2}\} - \{A_{2},A_{1}\}\right)$ and the Poisson bracket is 
\begin{align}
\{A_{1},A_{2}\} & = \sum_{i = 1}^{N} \left(\frac{\partial A_{1}}{\partial x_{i}}\frac{\partial A_{2}}{\partial p_{i}} - \frac{\partial A_{1}}{\partial p_{i}}\frac{\partial A_{2}}{\partial x_{i}}\right).
\end{align}
Overall, then, Eq.\eqref{eq: zeroth tier HEOM Wigner transform} becomes 
\begin{align}
\frac{\partial}{\partial t}\rho^{\text{el}}(\mathbf{x},\mathbf{p} ; t) & = \left\{H^{\text{el}}_{\text{S}}(\mathbf{x},\mathbf{p}),\rho^{\text{el}}(\mathbf{x},\mathbf{p} ; t) \right\}_{a}  \nonumber \\
& \:\:\:\:\: - i\left[H^{\text{el}}_{\text{S}}(\mathbf{x},\mathbf{p}),\rho^{\text{el}}(\mathbf{x},\mathbf{p} ; t)\right] \nonumber \\
& \:\:\:\:\: - i\sum_{j} \mathcal{A}^{\bar{\sigma}}_{\alpha,m} \rho^{(1),\text{el}}_{j}(\mathbf{x},\mathbf{p} ; t). \label{eq: zeroth tier HEOM Wigner transform final}
\end{align}
Each of the coupled equations in the hierarchy is then partially Wigner transformed, yielding in general
\begin{align}
 \frac{\partial}{\partial t}\rho^{(n),\text{el}}_{\mathbf{j}}(\mathbf{x},\mathbf{p} ; t) &  = \left\{H^{\text{el}}_{\text{S}}(\mathbf{x},\mathbf{p}),\rho^{(n),\text{el}}_{\mathbf{j}} \right\}_{a} \nonumber \\ 
& \:\:\:\:\:\:\: - i\left[H^{\text{el}}_{\text{S}}(\bold{x}),\rho^{(n),\text{el}}_{\mathbf{j}}\right]  - \left(\sum_{r = 1}^{N} \kappa^{}_{j_{r}}\right)\rho^{(n)}_{\mathbf{j}} \nonumber \\
& \:\:\:\:\:\:\:  - i\sum_{r=1}^{n} (-1)^{n-r}\mathcal{C}_{j_{r}}\rho^{(n-1),\text{el}}_{\mathbf{j}^{-}} \nonumber \\
& \:\:\:\:\:\:\:  - i\sum_{j} \mathcal{A}^{\bar{\sigma}}_{\alpha,m} \rho^{(n+1),\text{el}}_{\mathbf{j}^{+}}. \label{eq: nth tier HEOM Wigner transform}
\end{align}

In Eq.\eqref{eq: nth tier HEOM Wigner transform}, the partial Wigner transform of the part connecting the $n$th-tier ADO to the $(n+1)$th- and $(n-1)$th-tier ADOs only affects the ADOs themselves, as the superoperators $\mathcal{A}$ and $\mathcal{C}$ do not depend on the vibrational coordinates. It is also noted that, although the momenta enter $H^{\text{el}}_{S}(\mathbf{x},\mathbf{p})$ through the kinetic energies of the nuclei, they disappear in the commutator with the electronic density matrix. This means that the HEOM part of Eq.\eqref{eq: nth tier HEOM Wigner transform}, that is, everything except for the Poisson bracket, depends only on the vibrational coordinates, $\mathbf{x}$.

The reduced density matrix and all ADOs are next unfolded into vector format and collected into one joint vector, $\boldsymbol{\tilde{\rho}}^{\text{el}}(\mathbf{x},\mathbf{p} ; t) = \left(\rho^{\text{el}},\rho^{(1),\text{el}}_{\mathbf{j}},\dots,\rho^{(N),\text{el}}_{\mathbf{j}}\right)$. Everything can then be rewritten in the joint Liouville space of all the ADOs,
\begin{align}
\frac{\partial}{\partial t}\boldsymbol{\tilde{\rho}}^{\text{el}}(\mathbf{x},\mathbf{p} ; t) & = \{\!\!\{H^{\text{el}}_{\text{S}}(\mathbf{x},\mathbf{p}),\boldsymbol{\tilde{\rho}}^{\text{el}}(\mathbf{x},\mathbf{p} ; t) \}\!\!\}_{a} \nonumber \\
& \:\:\:\:\:\:\: - \mathcal{L}^{\text{el}}(\mathbf{x})\boldsymbol{\tilde{\rho}}^{\text{el}}(\mathbf{x},\mathbf{p} ; t). \label{eq: QCLE final}
\end{align}
The first term in Eq.\eqref{eq: QCLE final} contains the symmetrized Poisson bracket of the Wigner-transformed Hamiltonian with each Wigner-transformed ADO in $\boldsymbol{\tilde{\rho}}^{\text{el}}(\mathbf{x},\mathbf{p} ; t)$,
\begin{align}
& \{\!\!\{H^{\text{el}}_{\text{S}}(\mathbf{x},\mathbf{p}),\boldsymbol{\tilde{\rho}}^{\text{el}}(\mathbf{x},\mathbf{p} ; t) \}\!\!\}_{a} = \nonumber \\
& \:\:\:\:\: \left(\{H^{\text{el}}_{\text{S}},\rho^{(0),\text{el}}_{} \}_{a},\{H^{\text{el}}_{\text{S}},\rho^{(1),\text{el}}_{j_{1}} \}_{a},\{H^{\text{el}}_{\text{S}},\rho^{(1),\text{el}}_{j_{2}}.\}_{a},\dots\right),
\end{align}
which is then also unfolded into vector format. The second term, $\mathcal{L}^{\text{el}}(\mathbf{x})$, contains all dynamics of the electronic HEOM for a fixed vibrational frame, $\mathbf{x}$. Although not shown here, the same procedure can be applied to the BMME in Eq.\eqref{eq: BMME final} \cite{Dou2016a}.

The procedure above involves in general a classical approximation for the vibrational dynamics. If the electron-vibration coupling is linear in $\mathbf{x}$ and the vibrational potential is harmonic, as it is in all transport scenarios considered in this work, then the expansion in Eq.\eqref{eq: Wigner transform product expansion} naturally terminates after the first term and Eq.\eqref{eq: Wigner transform commutator approximation} is exact. For such a system, therefore, the mixed quantum-classical approach is just a formally exact rewriting of the original HEOM. In the next section, a further assumption will be introduced that will make the resulting equations of motion approximate, but for now, they remain formally exact. 

\subsection{Fokker-Planck equation}

The goal of this section is to derive a Fokker-Planck equation from Eq.\eqref{eq: QCLE final}. The approach is very similar to Ref. \cite{Dou2016a}, which is from the perspective of Born-Markov theory, but requires minor changes to suit HEOM. 

First, the steady state solution, $\boldsymbol{\tilde{\sigma}}^{\text{el}}_{\text{ss}}(\mathbf{x})$, is defined as the joint vector of all ADOs at vibrational frame $\mathbf{x}$ that satisfies $\mathcal{L}^{\text{el}}(\mathbf{x})\boldsymbol{\tilde{\sigma}}^{\text{el}}_{\text{ss}}(\mathbf{x}) = 0$, which is subject to the normalization condition that $\text{Tr}_{\text{el}}\left[\boldsymbol{\tilde{\sigma}}^{\text{el}}_{\text{ss}}(\mathbf{x})\right]  = 1$. Here, the electronic trace is an operation that returns the trace of only the reduced density matrix. Its action on the joint Wigner transform of all ADOs, for example, is
\begin{align}
\text{Tr}_{\text{el}}\left[\boldsymbol{\tilde{\rho}}^{\text{el}}(\mathbf{x},\mathbf{p} ; t) \right] & = \text{Tr}_{\text{el}}\left[\rho^{(0),\text{el}}(\mathbf{x},\mathbf{p} ; t) \right]  = \text{Tr}_{\text{el}}\left[\rho^{\text{el}}(\mathbf{x},\mathbf{p} ; t) \right].
\end{align} 
Next, the phase-space quasi-probability distribution is defined as
\begin{align}
A(\mathbf{x},\mathbf{p} ; t) & = \text{Tr}_{\text{el}}\left[\boldsymbol{\tilde{\rho}}^{\text{el}}(\mathbf{x},\mathbf{p} ; t)\right] = \text{Tr}_{\text{el}}\left[\rho^{\text{el}}(\mathbf{x},\mathbf{p} ; t) \right] ,
\end{align}
for which a Fokker-Planck equation will be derived.

The key step is to write $\boldsymbol{\tilde{\rho}}^{\text{el}}(\mathbf{x},\mathbf{p} ; t)$ as the probability to be at the steady state for a fixed vibrational frame, plus some difference $\mathbf{\tilde{B}}^{\text{el}}(\mathbf{x},\mathbf{p} ; t)$,
\begin{align}
\boldsymbol{\tilde{\rho}}^{\text{el}}(\mathbf{x},\mathbf{p} ; t) & = A(\mathbf{x},\mathbf{p} ; t)\boldsymbol{\tilde{\sigma}}^{\text{el}}_{\text{ss}}(\mathbf{x}) + \mathbf{\tilde{B}}^{\text{el}}(\mathbf{x},\mathbf{p} ; t), \label{eq: MFP starting point}
\end{align}
and consider the implications if $\mathbf{\tilde{B}}^{\text{el}}(\mathbf{x},\mathbf{p} ; t)$ is small.

Eq.\eqref{eq: MFP starting point} and Eq.\eqref{eq: QCLE final} are then used to write equations of motion for $A(\mathbf{x},\mathbf{p} ; t)$ and $\mathbf{\tilde{B}}^{\text{el}}(\mathbf{x},\mathbf{p} ; t)$. Under the Markovian assumption that the electronic degrees of freedom relax to equilibrium immediately for each vibrational coordinate, $\mathbf{x}$, the equation of motion for $\mathbf{\tilde{B}}^{\text{el}}(\mathbf{x},\mathbf{p} ; t)$ can be solved and substituted into that of $A(\mathbf{x},\mathbf{p} ; t)$ \cite{Dou2016a}. This results in a Markovian Fokker-Planck equation,
\begin{align}
\frac{\partial A(\mathbf{x},\mathbf{p} ; t)}{\partial t} & = - \sum_{i}\frac{p_{i}}{m_{i}} \frac{\partial A}{\partial x_{i}} + \sum_{ij}\gamma_{ij}(\mathbf{x}) \frac{\partial}{\partial p_{i}}\left(\frac{p_{j}}{m_{j}}A\right) \nonumber \\
& \:\:\:\:\:\: + \sum_{i}F^{}_{i}(\mathbf{x})\frac{\partial A}{\partial p_{i}}   + \sum_{ij}D_{ij}(\mathbf{x}) \frac{\partial^{2}A}{\partial p_{i}\partial p_{j}}, \label{eq: FP final}
\end{align}
with details found in Appendix \ref{app: On the derivation of the Markovian Fokker-Planck equation}. Here, the average electronic force is 
\begin{align}
F^{}_{i}(\mathbf{x}) & = - \text{Tr}_{\text{el}}\left[\frac{\partial H^{\text{el}}_{\text{S}}(\mathbf{x},\mathbf{p})}{\partial x_{i}}\boldsymbol{\tilde{\sigma}}^{\text{el}}_{\text{ss}}(\mathbf{x})\right], \label{eq: average electronic force HEOM 1}
\end{align}
the Markovian electronic friction tensor is 
\begin{align}
\gamma_{ij}(\mathbf{x}) & = -\lim_{\eta \rightarrow 0^{+}} \int^{\infty}_{0} dt \: \text{Tr}_{\text{el}}\left[\frac{\partial H^{\text{el}}_{\text{S}}}{\partial x_{i}}e^{-(\mathcal{L}^{\text{el}}(\mathbf{x}) + \eta)t}\frac{\partial \boldsymbol{\tilde{\sigma}}^{\text{el}}_{\text{ss}}}{\partial x_{j}}\right], \label{eq: friction calculation}
\end{align}
and the Markovian diffusion tensor is
\begin{align}
D_{ij}(\mathbf{x}) & = \frac{1}{2}\lim_{\eta \rightarrow 0^{+}} \int^{\infty}_{0} dt \: \text{Tr}_{\text{el}}\Big[\delta F_{i}(\mathbf{x})e^{-(\mathcal{L}^{\text{el}}(\mathbf{x}) + \eta)t}  \nonumber \\
& \:\:\:\:\: \hspace{2cm} \left(\delta F_{j}(\mathbf{x})\boldsymbol{\tilde{\sigma}}^{\text{el}}_{\text{ss}}(\mathbf{x}) + \boldsymbol{\tilde{\sigma}}^{\text{el}}_{\text{ss}}(\mathbf{x})\delta F_{j}(\mathbf{x})\right)\Big].\label{eq: diffusion calculation HEOM}
\end{align}
In Eq.\eqref{eq: diffusion calculation HEOM}, the operator containing the difference between the electronic force and the average electronic force has been introduced,
\begin{align}
\delta F_{j}(\mathbf{x}) & = \frac{\partial H^{\text{el}}_{\text{S}}}{\partial x_{i}} - \text{Tr}_{\text{el}}\left[\frac{\partial H^{\text{el}}_{\text{S}}}{\partial x_{i}}\boldsymbol{\tilde{\sigma}}^{\text{el}}_{\text{ss}}(\mathbf{x})\right].
\end{align}
Note that the Markovian assumption used to derive Eq.\eqref{eq: FP final} and the corresponding current-induced forces is completely separate to the assumption made in deriving the BMME in Sec. \ref{subsubsec: Born-Markov rate equation}, in which the bath electronic degrees of freedom relaxed quickly with respect to the system degrees of freedom. 

%

The Markovian Langevin equation for the $i^{\text{th}}$ vibrational coordinate, which corresponds to the Fokker-Planck equation in Eq.\eqref{eq: FP final}, is 
\begin{align}
m_{i}\ddot{x}_{i} & =  F^{}_{i}(\mathbf{x}) - \sum_{j} \gamma_{ij}(\mathbf{x}) \dot{x}_{j} + f_{i}(t), \label{eq: Langevin Markovian final}
\end{align}
where $f_{i}(t)$ is a Gaussian random force with white noise,
\begin{align}
\langle f_{i}(t)f_{j}(t')\rangle & = 2 D_{ij}(\mathbf{x}) \delta(t - t'). \label{eq: correlation function Markovian}
\end{align}

\subsection{Comments on the current-induced forces}

In this section, the important properties of the current-induced forces will be discussed, as well as relevant comments on their derivation and computation.

The form of the friction in Eq.\eqref{eq: friction calculation} is very similar to the expressions derived in Refs. \cite{Dou2017a,Dou2017c,Chen2019a}. These approaches, however, use the total electronic density matrix and the full time-evolution due to some general electronic system and do not explicitly include the splitting of electronic degrees of freedom into system and electrodes. A natural question, then, is whether the current-induced forces in Eqs.\eqref{eq: average electronic force HEOM 1}-\eqref{eq: diffusion calculation HEOM} are equivalent to those in Refs. \cite{Dou2017a,Dou2017c,Chen2019a}?

In Appendix \ref{app: Equivalence of the two friction tensors}, this question is answered and it is shown that the current-induced forces derived from the HEOM approach are, in fact, exactly the same quantities as have previously been reported in Refs. \cite{Dou2017a,Dou2017c,Chen2019a}. The only condition for this is that the vibrational degrees of freedom are restricted to the reduced system, an assumption that already underpins the mixed quantum-classical results of the previous section. The HEOM form of the current-induced forces, however, is an extremely useful representation, as it now provides access to a wide variety of transport scenarios, including systems with strong molecule-lead couplings and strong intra-system interactions, both in- and out-of-equilibrium.

In Refs. \cite{Dou2017a,Dou2017c,Chen2019a}, furthermore, several important properties of the current-induced forces were also derived. First, the friction and diffusion tensors are positive definite and symmetric. Second, the fluctuation-dissipation theorem is satisfied at equilibrium,
\begin{align}
D_{ij}(\mathbf{x}) = k_{B}T\gamma_{ij}(\mathbf{x}).
\end{align}
Finally, it was also shown that, under these conditions, the renormalized vibrational force is conservative, enabling one to define an effective potential of the mean force, $U_{\text{pmf}}(\mathbf{x})$, as 
\begin{align}
U_{\text{pmf}}(\mathbf{x}) & = \sum_{i} \int^{\mathbf{x}}_{\mathbf{x}_{0}} \: F^{}_{i}(\mathbf{x}') d\mathbf{x}'. \label{eq: potential of the mean force}
\end{align}
Given that the current-induced forces in Eqs.\eqref{eq: average electronic force HEOM 1}-\eqref{eq: diffusion calculation HEOM} are equivalent to those in Refs. \cite{Dou2017a,Dou2017c,Chen2019a}, they must also satisfy these properties.

Because a Markovian approximation has been made in deriving Eq.\eqref{eq: Langevin Markovian final}, such that the electronic degrees of freedom are assumed to relax immediately to the steady state for each vibrational frame, the friction and diffusion tensors depend only on the vibrational coordinates at each time, $t$, and not on the full history of the trajectory. If the explicit Markovian assumption is removed, such that there was still a separation of timescales between electronic and vibrational degrees of freedom, but $\frac{\partial \mathbf{\tilde{B}}^{\text{el}}}{\partial t} \neq 0$, then the derivation would yield a friction tensor with a simple form of memory; details are given in Appendix \ref{app: Non-Markovian friction tensor}. To include a the full non-Markovianity of the friction tensor, however, would require an approach such as that outlined in Ref. \cite{Chen2019a}. 

The friction tensor in Eq.\eqref{eq: friction calculation} is also conceptually similar to the one derived from Born-Markov theory in Ref. \cite{Dou2016a},
\begin{align}
\gamma_{ij}^{\text{BM}}(\mathbf{x}) & = -\lim_{\eta \rightarrow 0^{+}} \int^{\infty}_{0} dt \: \text{Tr}_{\text{el}}\left[\frac{\partial H^{\text{el}}_{\text{S}}}{\partial x_{i}}e^{-(\mathcal{L}^{\text{el}}_{\text{BM}}(\mathbf{x}) + \eta)t}\frac{\partial \sigma^{\text{el}}_{\text{ss}}}{\partial x_{j}}\right]. \label{eq: friction calculation BMME}
\end{align}
In Eq.\eqref{eq: friction calculation BMME}, however, the spatial derivative acts only on the steady state reduced system density matrix, $\sigma^{\text{el}}_{\text{ss}}(\mathbf{x})$. Similarly, the time-evolution defined in the $\mathcal{L}^{\text{el}}_{\text{BM}}(\mathbf{x})$ superoperator is also due to Born-Markov theory, and not the numerically exact HEOM method. 

So far the theory has been developed under the assumption that the vibrational degrees of freedom are restricted to the reduced nanosystem. There are many interesting systems and transport scenarios, such as current-induced bond rupture \cite{Ke2020}, where position-dependent molecular-lead couplings are necessary for a complete description. Although such position-dependent molecule-lead couplings can be treated within the HEOM formalism \cite{Ke2020,Ke2022}, they would substantially complicate the theory of electronic friction considered here. Eq.\eqref{eq: nth tier HEOM Wigner transform}, for example, would contain not only a Poisson bracket from the partial Wigner transform of $H_{S}\rho^{(n)}_{\mathbf{j}}$, but also from the $\mathcal{C}_{j_{r}}(\mathbf{x})\rho^{(n-1)}_{\mathbf{j}^{-}}$ and $\mathcal{A}^{\bar{\sigma}}_{\alpha,m}(\mathbf{x})\rho^{(n+1)}_{\mathbf{j}^{+}}$ terms, as these superoperators now also depend on $\mathbf{x}$. Similar to Ref.\cite{Dou2017b}, then, the friction would contain additional terms arising from these extra contributions.

Finally, although the Langevin equation in Eq.\eqref{eq: Langevin Markovian final} relates the vibrational forces linearly to the vibrational velocities, it includes the full non-linear dependence of the vibrational coordinates. This is different to several other treatments of the friction \cite{Hussein2010,Metelmann2011,Chen2019a,Chen2019b,Lue2012,Lue2010,Lue2011}, in which an expansion is made in the electron-vibration coupling, or, equivalently, small vibrational displacements are assumed.

\section{Results}\label{sec: Results}

In order to assess the performance of the HEOM approach when calculating current-induced forces nanosystems out-of-equilibrium, in this section, the electronic friction of three model systems is analyzed. 

In all calculations, the Fermi energy of the electrodes is set to $\epsilon_{F} = 0\text{ eV}$. 

\subsection{One electronic level coupled to one vibrational mode}\label{subsec: Resonant level}

\begin{figure}
\begin{centering}
\includegraphics[width = \columnwidth]{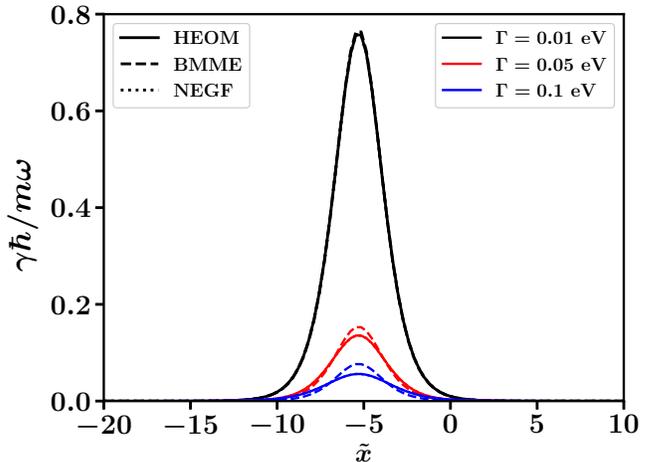}
\caption{Electronic friction as a function of the vibrational coordinate $\tilde{x}$ and for three different values of the molecule-lead coupling strength, $\Gamma$. The results are calculated at equilibrium, $\Phi = 0\text{ V}$, and the other parameters are chosen to match Ref. \cite{Dou2018a}: $\varepsilon_{0} = 0.15\text{ eV}$, $g = 0.02 \text{ eV}$, $k_{B}T = 0.026 \text{ eV}$, and $\hbar\omega = 0.003\text{ eV}$. Note that there is no regime in which exists a distinguishable difference between the exact friction and that calculated from the HEOM approach.}
\label{fig: 1}
\end{centering}
\end{figure}

The first, well-studied system is a single electronic level linearly coupled to a single vibrational coordinate. Using the dimensionless coordinate, $\tilde{x} = x\sqrt{m\omega/\hbar}$, and momentum, $\tilde{p} = p/\sqrt{m\omega\hbar}$, the corresponding Hamiltonian is 
\begin{align}
H^{\text{el}}_{\text{S}}(\tilde{x},\tilde{p}) & = \varepsilon(\tilde{x}) d^{\dag}d^{} + \frac{1}{2}\omega\left(\tilde{p}^{2} + \tilde{x}^{2}\right), \label{eq: Hamiltonian SRL}
\end{align}
where $\varepsilon(\tilde{x}) = \varepsilon_{0} + g\sqrt{2} \tilde{x}$. The vibrational degree of freedom moves according to a harmonic potential, $U_{0}(\tilde{x}) = \frac{1}{2}\omega \tilde{x}^{2}$, when the electronic level is unoccupied, and a shifted harmonic potential, $U_{1}(\tilde{x}) = U_{0}(\tilde{x}) + \varepsilon(\tilde{x})$, when occupied. 

Under the wide-band limit and because the system without the vibrational mode is non-interacting, the Markovian electronic friction tensor under a timescale separation assumption can be analytically derived from NEGFs, as in Refs. \cite{Dou2018a} and \cite{Preston2022},
 \begin{align}
\gamma(\tilde{x}) & = \frac{\Gamma}{2k_{B}T}\frac{m\omega}{\hbar}\left(\frac{\partial \varepsilon(\tilde{x})}{\partial \tilde{x}}\right)^{2}  \nonumber \\
& \: \times \int \frac{d\omega}{4\pi}\frac{\Gamma_{L}f^{+}_{L}(\omega)f^{-}_{L}(\omega) + \Gamma_{R}f^{+}_{R}(\omega)f^{-}_{R}(\omega)}{\left(\left[\omega - \varepsilon(\tilde{x})\right]^{2} + [\Gamma/2]^{2}\right)^{2}}, \label{eq: exact srl friction}
\end{align}
with $\Gamma = (\Gamma_{L} + \Gamma_{R})$. In Fig. \ref{fig: 1}, the analytic equilibrium friction in Eq.\eqref{eq: exact srl friction} is compared to the friction calculated from the two transport methods presented in Sec. \ref{sec: Quantum transport theory} and for three different molecule-lead couplings. For this system, the only feature of the electronic friction is a peak at $\tilde{x}_{0} = -\varepsilon_{0}/g\sqrt{2}$. This is the vibrational coordinate for which  $\varepsilon(\tilde{x})$ crosses the chemical potential of both electrodes and electron-hole pair creation is the dominant relaxation process, corresponding to the well-known mechanism of electronic friction \cite{Dou2018a}. One can see either from Fig. \ref{fig: 1} or Eq.\eqref{eq: exact srl friction} that the strength of the electronic friction is inversely related to the strength of the molecule-lead coupling. In this regime of slow vibrational motion, an increased $\Gamma$ means that the electrons do not spend long enough on the molecule to interact with the vibrational mode, so that both the damping rate and rate of energy injection are reduced.

Fig. \ref{fig: 1} also demonstrates the accuracy of the HEOM approach. For such a non-interacting electronic system, it has previously been shown that the corresponding hierarchical equations of motion terminates exactly at $2$nd-tier \cite{Zheng2008}, such that the friction here is equivalent to the NEGF result in Eq.\eqref{eq: exact srl friction}. The Born-Markov approach, in contrast, proves to have a limited regime of validity. When $\Gamma = 10\text{ meV}$, corresponding to the black lines in Fig. \ref{fig: 1}, all three approaches agree; as expected, Born-Markov theory performs well when $\Gamma \ll k_{B}T$. As the molecule-lead coupling increases to $\Gamma \approx 2k_{B}T$ and $\Gamma \approx 4k_{B}T$, corresponding to the red and blue lines, respectively, Born-Markov theory deviates from the exact analytic and HEOM results, producing a peak that is narrower and taller due to its incorrect treatment of level-broadening effects \cite{Dou2018a}.

\begin{figure}
\begin{centering}
\includegraphics[width = \columnwidth]{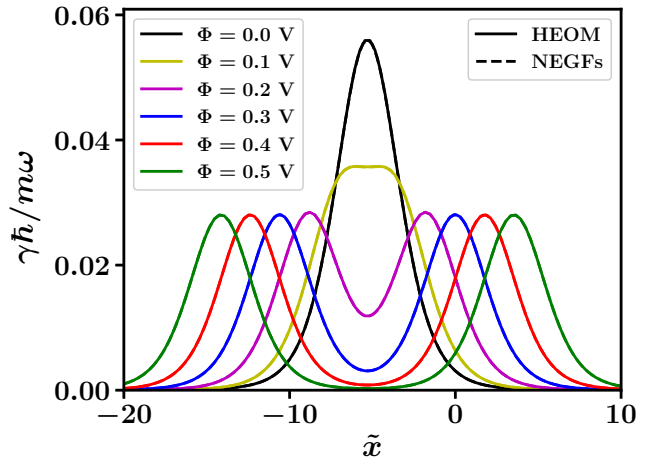}
\caption{Electronic friction as a function of the vibrational coordinate $\tilde{x}$ and at non-zero bias voltages. Apart from the voltage, which is applied symmetrically such that $\mu_{S} = -\mu_{D} = \Phi/2$, all parameters are the same as in Fig. \ref{fig: 1}. The molecule-lead coupling is $\Gamma = 0.1\text{ eV}$. Note that, even for $\Phi \neq 0$, there is no regime in which exists a distinguishable difference between the exact friction and that calculated from the HEOM approach.}
\label{fig: 2}
\end{centering}
\end{figure}

The HEOM approach works equally well for nonequilibrium scenarios, which is shown in Fig. \ref{fig: 2}. At a finite bias voltage, the friction displays two peaks at $\tilde{x}_{\alpha} = -(\varepsilon_{0} - \mu_{\alpha})/g\sqrt{2}$, where the renormalized energy level crosses the chemical potential of each electrode. The magnitude is also halved, as now there are only electron-hole pair creation processes associated with one electrode. Between the two peaks, $\gamma(\tilde{x})$ has a minimum at $\tilde{x}_{0}$, as the dominant transport process at this point is resonant tunneling, which leads to the well-understood mechanism of current-induced heating \cite{Erpenbeck2020,Haertle2011a}. 

\subsection{Anderson impurity coupled to one vibrational mode}\label{subsec: Anderson impurity}

In this section, a model describing an Anderson impurity coupled to one vibrational mode is considered, which, due to the presence of electron-electron interactions, cannot be solved exactly. The system Hamiltonian now contains a spin-degenerate level linearly coupled to the vibrational coordinate of a harmonic vibrational mode, with a non-zero, coordinate-independent Coulomb repulsion between electrons of opposite spin,
\begin{align}
H^{\text{el}}_{\text{S}}(\tilde{x},\tilde{p}) & = \varepsilon(\tilde{x}) \sum_{s \in \{\uparrow,\downarrow\}}d^{\dag}_{s}d^{}_{s} + U d^{\dag}_{\uparrow}d^{}_{\uparrow}d^{\dag}_{\downarrow}d^{}_{\downarrow} \nonumber \\ 
& \:\:\:\:\:\: + \frac{1}{2}\omega\left(\tilde{p}^{2} + \tilde{x}^{2}\right),
\end{align}
with the bath annihilation and creation operators also now carrying a spin index,
\begin{align}
H_{\text{B}} & = \sum_{\alpha \in \{L,R\}}\sum_{k_{\alpha},s} \varepsilon_{k_{\alpha},s} c^{\dag}_{k_{\alpha},s}c^{}_{k_{\alpha},s} \\
H_{\text{SB}} & = \sum_{\alpha,k_{\alpha}}\sum_{s} V_{k_{\alpha},s}\left(c^{\dag}_{k_{\alpha},s}d^{}_{s} + d^{\dag}_{s}c^{}_{k_{\alpha},s}\right).
\end{align}

\begin{figure}
\begin{centering}
\includegraphics[width = \columnwidth]{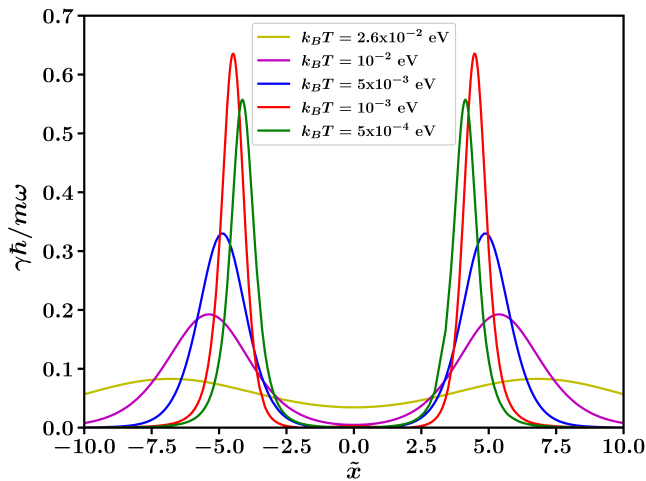}
\caption{Electronic friction as a function of vibrational coordinate $\tilde{x}$ for an Anderson impurity and at a range of temperatures. Calculations are performed at equilibrium: $\Phi = 0 \text{ V}$. To compare directly with Ref. \cite{Dou2017c}, parameters were chosen as $\varepsilon_{0} = -0.05\text{ eV}$, $U = 0.1\text{ eV}$, $g = 0.0075\text{ eV}$, and $ \Gamma = 0.01\text{ eV}$.}
\label{fig: 3}
\end{centering}
\end{figure}

As with the single electronic level model, the Anderson impurity model has already been investigated in the context of electronic friction \cite{Dou2017c,Askerka2016,Dzhioev2013,Plihal1999,Plihal1998}. In Ref. \cite{Dou2017c}, for example, the authors diagonalize the electronic part of the total Hamiltonian with a numerical renormalization group (NRG) technique and then use NEGFs to calculate $\gamma(\tilde{x})$ at equilibrium, demonstrating the superiority of NRG to a mean-field theory approach \cite{Askerka2016,Dzhioev2013}, which does not correctly recover the doubly-peaked behavior of $\gamma(\tilde{x})$. The HEOM method developed in this paper, in comparison, is able to perform a numerically exact calculation of the friction for this model out-of-equilibrium.  

To compare with the NEGF approach, Fig. \ref{fig: 3} shows, for a range of temperatures, the electronic friction as a function of the vibrational coordinate for the same parameters as in Fig. 1a from Ref. \cite{Dou2017c}. At equilibrium, the friction displays two peaks around the points where there is an electron-hole pair creation resonance, $\varepsilon(\tilde{x}) = 0$ and $\varepsilon(\tilde{x}) + U = 0$. At high to moderate temperatures, the HEOM approach yields the same results as NEGFs. At low temperatures, however, so many poles are required for the Pad\'{e} decomposition of the Fermi-Dirac function that the HEOM become numerically intractable, a well-known problem that limits the temperature range of the method. Recent proposals \cite{Chen2022,Xu2022,Nakatsukasa2018} for a more efficient pole expansion appear promising in extending the HEOM approach to low temperatures. 

\begin{figure}
\begin{centering}
\includegraphics[width = \columnwidth]{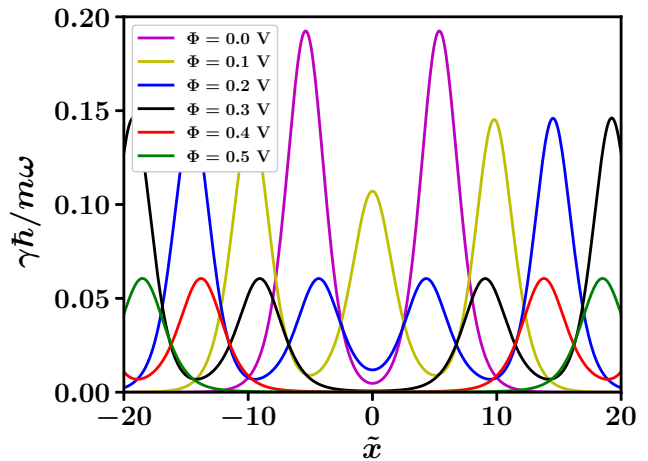}
\caption{Electronic friction as a function of vibrational coordinate $\tilde{x}$ for an Anderson impurity model and at a range of bias voltages. Calculations are performed for $k_{B}T = \Gamma = 0.01\text{ eV}$, and all other parameters are the same as in Fig. \ref{fig: 3}.}
\label{fig: 4}
\end{centering}
\end{figure}

At finite bias voltage, which is shown in Fig. \ref{fig: 4}, the two friction peaks split into four, as there is now a resonance whenever an electron-hole pair creation process is resonant with the chemical potential of an electrode: $\varepsilon(\tilde{x}) = \mu_{\alpha}$ and $\varepsilon(\tilde{x}) + U = \mu_{\alpha}$. The peaks at $\varepsilon(\tilde{x}) = \mu_{\alpha}$ are larger than those at $\varepsilon(\tilde{x}) + U = \mu_{\alpha}$, because in the first case the degenerate level reaches $\mu_{\alpha}$ and two electron-hole pair creation processes become available, whereas in the second it is only the doubly-occupied process that activates. At $\Phi = 0.1\text{ V}$, there is only one extra peak at $\tilde{x} = 0$, because $\varepsilon(0) + U = \mu_{S}$ and $\varepsilon(0) = \mu_{D}$.

\subsection{One electronic level coupled to one classical and one quantum vibrational mode}\label{subsec: High frequency mode}

\begin{figure}
\begin{subfigure}[b]{0.0\textwidth}
\phantomcaption\label{fig: 5a}
\end{subfigure}
\begin{subfigure}[b]{0.0\textwidth}
\phantomcaption\label{fig: 5b}
\end{subfigure}
\begin{centering}
\includegraphics[width = \columnwidth]{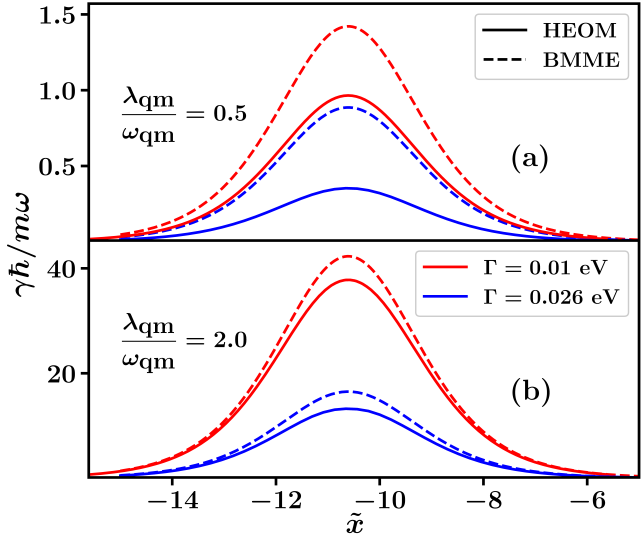}
\caption{Electronic friction in equilibrium as a function of vibrational coordinate $\tilde{x}$ for a single electronic level coupled to a high-frequency quantum vibrational mode and low-frequency classical vibrational mode. Results are shown for two different molecule-lead couplings, $\Gamma = 0.01\text{ eV}$ (red) and $\Gamma = 0.026\text{ eV}$ (blue), and from the HEOM and Born-Markov approaches, corresponding to the solid and dashed lines, respectively. Other parameters are $g_{\text{cl}} = 0.02\text{ eV}$, $k_{B}T = 0.026\text{ eV}$, $\hbar\omega_{\text{qm}} = 0.2\text{ eV}$, and in (a) $\lambda_{\text{qm}} = 0.1\text{ eV}$, while in (b) $\lambda_{\text{qm}} = 0.4\text{ eV}$. The energy, $\varepsilon_{0}$, is chosen so that the renormalized energy level after the small polaron transformation is always $\varepsilon_{0} - \frac{\lambda_{\text{qm}}^{2}}{\omega_{\text{qm}}} = 0.3\text{ eV}$.}
\label{fig: 5}
\end{centering}
\end{figure}

\begin{figure*}
\begin{centering}
\includegraphics[width = \textwidth]{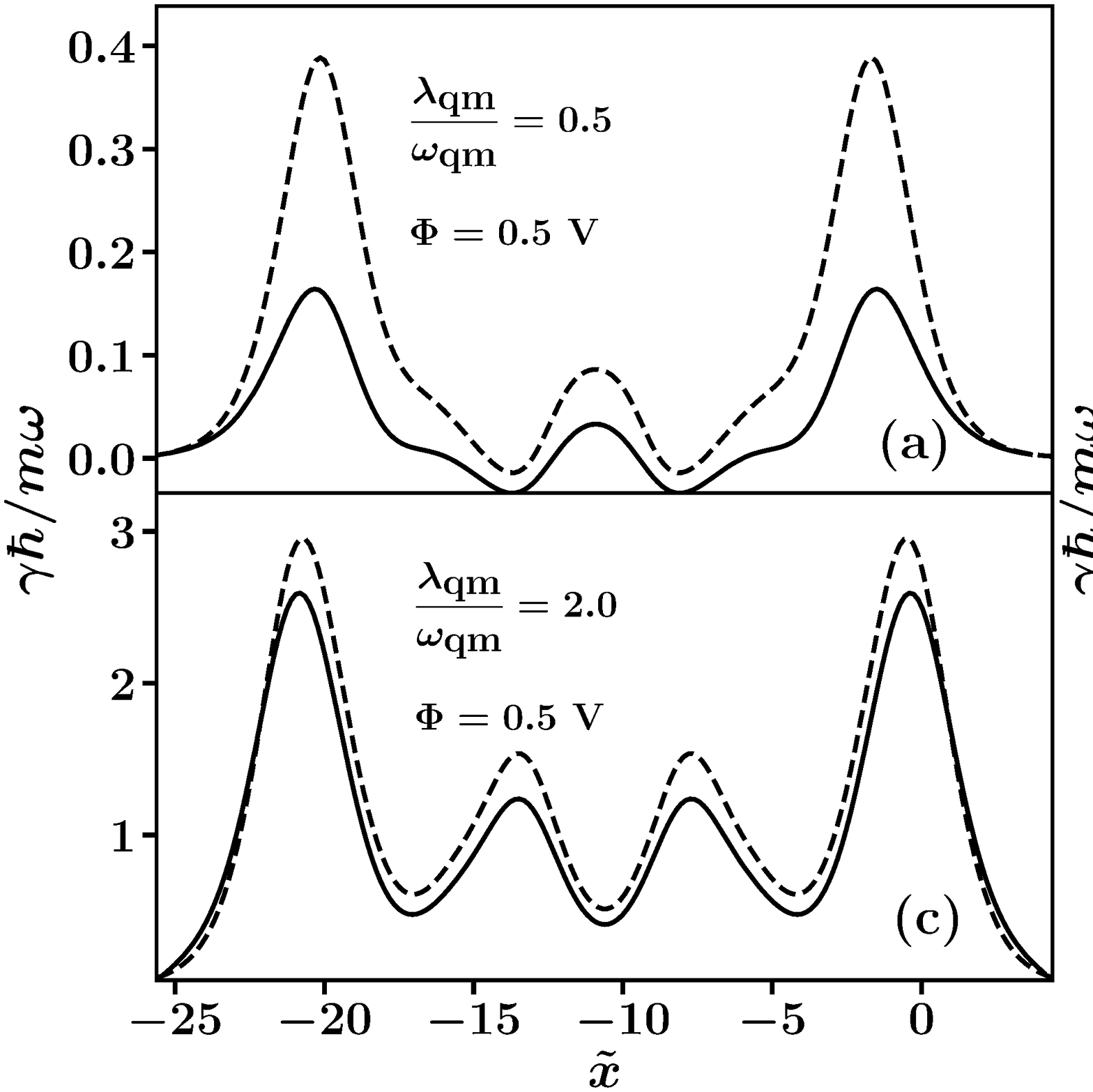}
\begin{subfigure}[b]{0.0\textwidth}
\phantomcaption\label{fig: 6a}
\end{subfigure}
\begin{subfigure}[b]{0.0\textwidth}
\phantomcaption\label{fig: 6b}
\end{subfigure}
\begin{subfigure}[b]{0.0\textwidth}
\phantomcaption\label{fig: 6c}
\end{subfigure}
\begin{subfigure}[b]{0.0\textwidth}
\phantomcaption\label{fig: 6d}
\end{subfigure}
\caption{Electronic friction at finite bias voltages and as a function of vibrational coordinate $\tilde{x}$ for an electronic level coupled to a high-frequency quantum vibrational mode as well as a low-frequency classical vibrational mode. The molecule-lead coupling is $\Gamma = 0.026\text{ eV}$. The upper row corresponds to $\lambda_{\text{qm}} = 0.1\text{ eV}$, while the lower row corresponds to $\lambda_{\text{qm}} = 0.4\text{ eV}$. The left column is held at a voltage of $\Phi = 0.5\text{ V}$, while the right column is held at a voltage of $\Phi = 1.0\text{ V}$. All other parameters are the same as in Fig. \ref{fig: 5}.}
\label{fig: 6}
\end{centering}
\end{figure*}

In the final example, a single electronic level is coupled to two vibrational modes. One, which is treated classically, is subject to the limit of slow vibrational driving, with a low frequency $\hbar\omega_{\text{cl}} \ll \Gamma $, and one, which remains in the fully quantum mechanical HEOM treatment, has a high frequency $\hbar\omega_{\text{qm}} > \Gamma $. Although this degree of freedom is vibrational and not electronic, it can still be included in $H^{\text{el}}_{\text{S}}(x_{\text{cl}},p_{\text{cl}})$ as long as it follows the timescale separation necessary for the adiabatic approximation. As discussed in Ref. \cite{Chen2019b}, in which the electronic friction for a similar model was calculated under different assumptions, such a model could describe a single-molecule in an optical cavity, where the high-frequency mode describes a cavity mode with a strong light-matter interaction \cite{Kongsuwan2018,Chikkaraddy2016}. An alternative experimental realization would be a molecule with two bonds of distinctly different strengths. 

Similar to the previous two case studies, the electronic level is linearly coupled to both the classical and quantum vibrational coordinates, such that the resulting Hamiltonian is
\begin{align}
H^{\text{el}}_{\text{S}}(\tilde{x}_{\text{cl}},\tilde{p}_{\text{cl}}) & = \varepsilon(\tilde{x}_{\text{cl}}) d^{\dag}d^{} + \omega_{\text{qm}}b^{\dag}b^{} + \lambda_{\text{qm}} (b^{\dag} + b^{})d^{\dag}d^{}  \nonumber \\ 
& \:\:\:\:\:\: + \frac{1}{2}\omega_{\text{cl}}\left(\tilde{p}_{\text{cl}}^{2} + \tilde{x}_{\text{cl}}^{2}\right).
\end{align}
The bosonic annihilation and creation operators are related to the dimensionless coordinate, $\hat{\tilde{x}}_{\text{qm}} = \hat{x}_{\text{qm}}\sqrt{m_{\text{qm}}\omega_{\text{qm}}/\hbar}$, and momentum, $\hat{\tilde{p}}_{\text{qm}} = \hat{p}_{\text{qm}}/\sqrt{m_{\text{qm}}\omega_{\text{qm}}\hbar}$, of the quantum vibrational mode via 
\begin{align}
b & = \frac{1}{\sqrt{2}}(\hat{\tilde{x}}_{\text{qm}} + i\hat{\tilde{p}}_{\text{qm}}) \hspace{0.25cm} \text{ and } \hspace{0.25cm} b^{\dag} = \frac{1}{\sqrt{2}}(\hat{\tilde{x}}_{\text{qm}} - i\hat{\tilde{p}}_{\text{qm}}).
\end{align}
Unlike the previous two case studies, however, it is difficult to investigate this model with other transport methods, as it contains a strong electron-phonon interaction. This model serves, therefore, as an excellent example of the new variety of nanosystems now accessible by semi-classical Langevin equations. 
%
%
%
The quantum part of the Hamiltonian can be diagonalized via a small polaron transformation \cite{Lang1963}, which yields
\begin{align}
H^{\text{el}}_{\text{S}}(\tilde{x}_{\text{cl}},\tilde{p}_{\text{cl}}) & = \tilde{\varepsilon}(\tilde{x}_{\text{cl}}) d^{\dag}d^{} + \omega_{\text{qm}}b^{\dag}b^{}  + \frac{1}{2}\omega_{\text{cl}}\left(\tilde{p}_{\text{cl}}^{2} + \tilde{x}_{\text{cl}}^{2}\right), \label{eq: Hamiltonian Holstein final}
\end{align}
a Hamiltonian similar to Eq.\eqref{eq: Hamiltonian SRL}, except that the electronic energy level has been renormalized to 
\begin{align}
\tilde{\varepsilon}(\tilde{x}_{\text{cl}}) & = \varepsilon_{0} - \frac{\lambda_{\text{qm}}^{2}}{\omega_{\text{qm}}} + \sqrt{2}g_{\text{cl}}\tilde{x}_{\text{cl}}.
\end{align} 
The quantum transport under the small polaron transformation occurs between eigenstates of the system Hamiltonian $|q,\nu\rangle \rightarrow |q',\nu'\rangle$, where the quantum numbers $q = 0,1$ and $\nu = 0,1,2,\dots$ refer to charge and phonon occupancy, respectively. Transporting electrons can exchange quanta of vibrational energy with the quantum harmonic oscillator, with the transition probability between vibrational states of the occupied and unoccupied system given by the overlap of the wave functions of the original and displaced quantum harmonic oscillator, which is the Franck-Condon matrix,
\begin{align}
X_{\nu,\nu'} & =  \langle \nu | X | \nu' \rangle = \langle \nu | e^{\frac{\lambda_{\text{qm}}}{\omega_{\text{qm}}}(b - b^{\dag})} | \nu' \rangle.
\end{align} 
In this transport picture, the ratio $\lambda_{\text{qm}}/\omega_{\text{qm}}$ determines the structure of the Franck-Condon matrix. For small $\lambda_{\text{qm}}$, the $X_{\nu,\nu'}$ are close to diagonal and the transitions between low-lying vibrational states are favored, while the opposite is true for strong $\lambda_{\text{qm}}$ \cite{Koch2005,Koch2006a,Haertle2011a,Haertle2011b}.

Due to the strong interaction between the electronic level and the quantum vibrational mode, Eq.\eqref{eq: Hamiltonian Holstein final} is a model well-suited to the HEOM approach. Under the appropriate high-temperature, weak-coupling limit, one could also use a Born-Markov master equation, as this formulation is also able to treat strong intra-system interactions exactly. To test this, Fig. \ref{fig: 5}\hyperref[fig: 5]{(a)} and Fig. \ref{fig: 5}\hyperref[fig: 5]{(b)} show the Born-Markov friction against the HEOM friction at equilibrium and for two different molecule-lead couplings, $\Gamma$, and electron-phonon couplings, $\lambda_{\text{qm}}$. As expected, in the strong-coupling regime of $\Gamma = k_{B}T = 0.026\text{ eV}$, Born-Markov theory does not reproduce the numerically exact HEOM result. Even in the weak-coupling regime of $\Gamma = 0.01\text{ eV} < k_{B}T$,  furthermore, there are significant differences between $\gamma$ calculated from the HEOM and BMME approaches, indicating that the electronic friction for this model requires the superior HEOM approach even for moderate temperatures and molecule-lead couplings. 

Apart from the differences between the Born-Markov and HEOM approaches, the qualitative friction behavior strongly resembles that of the single electronic level in equilibrium, in that there is a single peak at 
\begin{align}
\tilde{x}_{0} & = -\frac{\varepsilon_{0} - \lambda_{\text{qm}}^{2}/\omega_{\text{qm}}}{g_{\text{cl}}\sqrt{2}}.
\end{align}
In Fig. \ref{fig: 5}\hyperref[fig: 5]{(b)}, however, the height of the friction peak is an order of magnitude larger than it is in the non-interacting case, which originates from the additional electron-hole pair creation processes introduced by the quantum vibrational mode, as was discussed in Ref. \cite{Haertle2011a}. At equilibrium, an electron-hole pair can be created with an effective cooling process when an electron tunnels in from electrode $\alpha$ and absorbs one or more vibrational quanta, allowing it to then tunnel out to an unoccupied state in the same electrode; see Fig.(2) in Ref. \cite{Haertle2011a} for a schematic example. Because the transition probability from the vibrational ground to excited states, $|X_{0,\nu}|^{2} = \frac{1}{\nu !}(\frac{\lambda_{\text{qm}}}{\omega_{\text{qm}}})^{2\nu}e^{-(\lambda_{\text{qm}}/\omega_{\text{qm}})^{2}}$, is not suppressed for  $\lambda_{\text{qm}}/\omega_{\text{qm}} = 2$, these electron-hole pair creation processes contribute significantly to the friction. For the weaker electron-phonon coupling in Fig. \ref{fig: 5}\hyperref[fig: 5]{(a)}, where $\lambda_{\text{qm}}/\omega_{\text{qm}} = 0.5$, the transition probabilities are suppressed and the friction is also quantitatively similar to the non-interacting system in Fig. \ref{fig: 1}.

At finite bias voltages, the electronic friction displays a much richer structure, which is shown in Fig. \ref{fig: 6}. In Fig. \ref{fig: 6}\hyperref[fig: 6]{(c)} and Fig. \ref{fig: 6}\hyperref[fig: 6]{(d)}, where $\lambda_{\text{qm}}/\omega_{\text{qm}} = 2$, the friction displays characteristic peaks at points where 
\begin{align}
\tilde{x}_{\alpha,\nu} & = \frac{\mu_{\alpha} - (\varepsilon_{0} - \lambda_{\text{qm}}^{2}/\omega_{\text{qm}} + q\omega_{\text{qm}})}{g_{\text{cl}}\sqrt{2}}:
\end{align}
that is, points where the renormalized energy level plus some integer of vibrational quanta crosses the chemical potential of electrode $\alpha$. At each $\tilde{x}_{\alpha,\nu}$, the electron-hole pair creation process in electrode $\alpha$ associated with the absorption of $\nu$ vibrational quanta is enhanced. These side peaks are not visible at equilibrum, however, as only the ground to ground state transition is energetically allowed \cite{Koch2005}.

In contrast, the friction minima occur when the chemical potential of electrode $\alpha$ is located between two such quasi-levels and the system experiences current-induced heating. The peaks also decrease as the renormalized level is pushed further into the center of the bias window; the further $\varepsilon_{0} - \lambda_{\text{qm}}^{2}/\omega_{\text{qm}} + g_{\text{cl}}\sqrt{2}\tilde{x}$ is away from $\mu_{\alpha}$, the more vibrational quanta are required for electron-hole pair creation, and these processes are generally less probable. Finally, the main difference between Figs. \ref{fig: 6}\hyperref[fig: 6]{(c)} and Fig. \ref{fig: 6}\hyperref[fig: 6]{(d)} is that the higher voltage in Fig. \ref{fig: 6}\hyperref[fig: 6]{(d)} energetically allows more vibrational transitions to contribute to electron-hole pair creation, causing the extra two friction peaks. 

In Fig. \ref{fig: 6}\hyperref[fig: 6]{(a)} and Fig. \ref{fig: 6}\hyperref[fig: 6]{(b)}, the electron-phonon coupling is again weak, $\lambda_{\text{qm}}/\omega_{\text{qm}} = 0.5$, and electron-hole pair creation via absorption of vibrational quanta is suppressed with increasing voltage. At $\tilde{x}_{\alpha,0}$, the resonant electron-hole pair creation process still yields a friction peak because the vibrational ground to excited state transition has a strong Franck-Condon factor. At each of the other $\tilde{x}_{\alpha,\nu \neq 0}$, however, the electronic friction experiences a local or global minimum, as the $|X_{0,\nu>0}|^{2}$ are suppressed for increasing $\nu$ and $\lambda_{\text{qm}}/\omega_{\text{qm}} < 1$. In this regime, the electronic friction actually becomes negative. This implies that the loss of cooling by electron-hole pair creation not only heats the quantum vibrational mode, but that energy is also being pumped into the classical mode by the usually dissipative processes, driving it to instability. The identification of negative friction is of high interest in transport through nanostructures, as it provides an easy insight into regimes of instability where the junction can break \cite{Preston2022,Lue2012}. 

\section{Conclusion}\label{sec: Conclusion}

In this work, the HEOM approach to nonequilibrium charge transport through nanosystems was used to develop a new method for calculating current-induced forces. Similar to other master equation approaches to electronic friction \cite{Dou2016a}, the fully quantum HEOM is first transformed to a QCLE and then, under an adiabatic approximation for the vibrational degrees of freedom, to a Markovian Fokker-Planck equation, from which the average electronic force, Markovian friction tensor, Markovian diffusion tensor, and corresponding Langevin equation are identified. These quantities were then shown to be equivalent to previous derivations of the current-induced forces under a timescale separation, but in a more computationally useful form. The electronic friction, for example, can now be calculated for junctions with both strong molecule-lead couplings and strong intra-system interactions. The resulting friction and diffusion tensors, furthermore, are completely general in- and out-of-equilibrium.

The HEOM approach to calculating electronic friction was applied to three models of different complexity. First, a single non-interacting electronic level coupled linearly to one classical vibrational mode was considered, a system for which exact results can be calculated; these were used to demonstrate the numerical accuracy of the HEOM approach. In the second model, an Anderson impurity coupled to one classical vibrational mode was considered. Because adding extra levels and interactions does not complicate the HEOM theory, this system can be treated easily both in- and out-of-equilibrium. Finally, a system with an additional interaction between the electronic level and a quantum vibrational mode was considered, which yielded a richer structure in the friction of the classical mode. In particular, it was shown that for a simple one level, two mode system, heating of the quantum vibrational mode caused negative friction in the classical mode, indicating that it is also being driven to instability. 

\section{Acknowledgements}\label{sec: Acknowledgements}

This work was supported by the German Research Foundation (DFG) through FOR5099. S.L.R thanks the Alexander von Humboldt Foundation for the award of a Research Fellowship. Furthermore, support by the state of Baden-W\"{u}rttemberg through bwHPC and the DFG through Grant No. INST 40/575-1 FUGG (JUSTUS 2 cluster) is gratefully acknowledged.

\appendix 

\section{On the derivation of the Markovian Fokker-Planck equation}\label{app: On the derivation of the Markovian Fokker-Planck equation}

In this appendix, the Markovian Fokker-Planck equation in Eq.\eqref{eq: FP final} will be explicitly derived. First, the equations of motion for $A(\mathbf{x},\mathbf{p} ; t)$ and $\mathbf{\tilde{B}}^{\text{el}}(\mathbf{x},\mathbf{p} ; t)$ are 
\begin{align}
\frac{\partial A}{\partial t} & = \text{Tr}_{\text{el}}[\{\!\!\{H^{\text{el}}_{\text{S}},A\boldsymbol{\tilde{\sigma}}^{\text{el}}_{\text{ss}}\}\!\!\}_{a}] + \text{Tr}_{\text{el}}[\{\!\!\{H^{\text{el}}_{\text{S}},\mathbf{\tilde{B}}^{\text{el}}\}\!\!\}_{a}]
\nonumber \\
& = - \sum_{i} \frac{p_{i}}{m_{i}} \frac{\partial A}{\partial x_{i}} + \sum_{i} \text{Tr}_{\text{el}}\left[\frac{\partial H^{\text{el}}_{\text{S}}}{\partial x_{i}} \boldsymbol{\tilde{\sigma}}^{\text{el}}_{\text{ss}}\right]\frac{\partial A}{\partial p_{i}} \nonumber \\ 
& \:\:\:\:\:\: + \sum_{i} \text{Tr}_{\text{el}}\left[\frac{\partial H^{\text{el}}_{\text{S}}}{\partial x_{i}} \frac{\partial \mathbf{\tilde{B}}^{\text{el}}}{\partial p_{i}}\right] \hspace{1cm}\text{ and } \label{eq: dA/dt} \\
\frac{\partial \mathbf{\tilde{B}}^{\text{el}}}{\partial t} & = \{\!\!\{H^{\text{el}}_{\text{S}},\mathbf{\tilde{B}}^{\text{el}}\}\!\!\}_{a} - \boldsymbol{\tilde{\sigma}}^{\text{el}}_{\text{ss}}\text{Tr}_{\text{el}}[\{\!\!\{H^{\text{el}}_{\text{S}},\mathbf{\tilde{B}}^{\text{el}}\}\!\!\}_{a}] - \mathcal{L}^{\text{el}}\mathbf{\tilde{B}}^{\text{el}} \nonumber \\ 
& \:\:\:\:\:\: + \{\!\!\{H^{\text{el}}_{\text{S}},A\boldsymbol{\tilde{\sigma}}^{\text{el}}_{\text{ss}}\}\!\!\}_{a} - \boldsymbol{\tilde{\sigma}}^{\text{el}}_{\text{ss}}\text{Tr}_{\text{el}}[\{\!\!\{H^{\text{el}}_{\text{S}},A\boldsymbol{\tilde{\sigma}}^{\text{el}}_{\text{ss}}\}\!\!\}_{a}].
\label{eq: dB/dt}
\end{align}
where the $(\mathbf{x},\mathbf{p} ; t)$ notation has been suppressed for brevity and $\text{Tr}_{\text{el}}\left[\mathcal{L}^{\text{el}}\boldsymbol{\tilde{\rho}}^{\text{el}}\right] = 0$ has been used. This relation can be seen if one considers that $\text{Tr}_{\text{el}}\left[\dots\right]$ only traces over the $0$th-tier ADO, and the corresponding part of the hierarchy in $\mathcal{L}^{\text{el}}$ contains only commutators,
\begin{align}
\text{Tr}_{\text{el}}\left[\mathcal{L}^{\text{el}}\boldsymbol{\tilde{\rho}}^{\text{el}}\right] & =  -i\text{Tr}_{\text{el}}\left(\left[H^{\text{el}}_{\text{S}},\rho^{\text{el}}\right]\right) \nonumber \\ 
& \:\:\:\: - i\sum_{j} V_{\alpha,m} \text{Tr}_{\text{el}}\left(\left[d^{\bar{\sigma}},\boldsymbol{\tilde{\rho}}^{(1),\text{el}}_{j}\right]\right) \\
& = 0.
\end{align}

The first few terms in Eq.\eqref{eq: dB/dt} involve derivatives of $\mathbf{\tilde{B}}^{\text{el}}$ with respect to $t$, $x_{i}$, and $p_{i}$, due to the Poisson bracket. As has been discussed in Refs. \cite{Dou2017c,Dou2015d}, the limit of slow vibrational degrees of freedom in comparison to fast electronic degrees of freedom allows these terms to be neglected, as $\mathbf{\tilde{B}}^{\text{el}}$ should be much smaller than $A$. Eq.\eqref{eq: dB/dt} becomes, therefore, 
\begin{align}
\mathcal{L}^{\text{el}}\mathbf{\tilde{B}}^{\text{el}} & = \{\!\!\{H^{\text{el}}_{\text{S}},A\boldsymbol{\tilde{\sigma}}^{\text{el}}_{\text{ss}}\}\!\!\}_{a} - \boldsymbol{\tilde{\sigma}}^{\text{el}}_{\text{ss}}\text{Tr}_{\text{el}}[\{\!\!\{H^{\text{el}}_{\text{S}},A\boldsymbol{\tilde{\sigma}}^{\text{el}}_{\text{ss}}\}\!\!\}_{a}] \label{eq: B TEV 2} \\
& = - \sum_{j} \frac{p_{j}}{m_{j}}\frac{\partial \boldsymbol{\tilde{\sigma}}^{\text{el}}_{\text{ss}}}{\partial x_{j}}A - \sum_{j} \text{Tr}_{\text{el}}\left[\frac{\partial H^{\text{el}}_{\text{S}}}{\partial x_{j}}\boldsymbol{\tilde{\sigma}}^{\text{el}}_{\text{ss}}\right]\frac{\partial A}{\partial p_{j}}\boldsymbol{\tilde{\sigma}}^{\text{el}}_{\text{ss}} \nonumber \\ 
& \:\:\:\:\:\: + \frac{1}{2} \sum_{j} \left(\frac{\partial H^{\text{el}}_{\text{S}}}{\partial x_{j}}\boldsymbol{\tilde{\sigma}}^{\text{el}}_{\text{ss}} + \boldsymbol{\tilde{\sigma}}^{\text{el}}_{\text{ss}}\frac{\partial H^{\text{el}}_{\text{S}}}{\partial x_{j}}\right)\frac{\partial A}{\partial p_{j}}. \label{eq: B TEV 3}
\end{align}
With the $(\mathbf{x},\mathbf{p} ; t)$ notation now explicitly written, and using the identity 
\begin{align}
[\mathcal{L}^{\text{el}}(\mathbf{x})]^{-1} & = \lim_{\eta \rightarrow 0^{+}} \int^{\infty}_{0} dt \: e^{-(\mathcal{L}^{\text{el}}(\mathbf{x}) + \eta)t},
\end{align}
$\mathcal{L}^{\text{el}}(\mathbf{x})$ is formally inverted and $\mathbf{\tilde{B}}^{\text{el}}(\mathbf{x},\mathbf{p} ; t)$ can be directly calculated. This is then substituted into the equation of motion for $A(\mathbf{x},\mathbf{p} ; t)$, yielding the Markovian Fokker-Planck equation in Eq.\eqref{eq: FP final}.

\section{Comparison to previous approaches}\label{app: Equivalence of the two friction tensors}

The goal of this section is to show that the current-induced forces derived via the HEOM approach in the main text are equivalent to other approaches that treat all electronic degrees of freedom without an explicit system-bath partitioning. See, for example, Eq.(11), Eq.(18), and Eq.(19) in Ref. \cite{Dou2018a} or the Markovian version of Eq.(27) in Ref. \cite{Chen2019a}. In these approaches, the average force is 
\begin{align}
F^{}_{i}(\mathbf{x}) & = - \text{Tr}_{\text{tot},\text{el}}\left[\frac{\partial H^{\text{el}}}{\partial x_{i}}\rho^{\text{el}}_{\text{tot},\text{ss}}\right], \label{eq: average force calculation full 1}
\end{align}
the Markovian electronic friction is 
\begin{align}
\gamma_{ij}(\mathbf{x}) & = -\int^{\infty}_{0} dt \: \text{Tr}_{\text{tot},\text{el}}\left[\frac{\partial H^{\text{el}}}{\partial x_{i}}e^{-iH^{\text{el}}t}\frac{\partial \rho^{\text{el}}_{\text{tot},\text{ss}}}{\partial x_{j}}e^{iH^{\text{el}}t}\right], \label{eq: friction calculation full 1}
\end{align}
and the Markovian diffusion tensor is 
\begin{align}
D_{ij}(\mathbf{x}) & = \int^{\infty}_{0} dt \: \text{Tr}_{\text{tot},\text{el}}\left[\delta F_{i}e^{-iH^{\text{el}}t}\left(\delta F_{j} \rho^{\text{el}}_{\text{tot},\text{ss}} \right.\right.\nonumber \\
& \hspace{2.7cm} \left.\left. +\rho^{\text{el}}_{\text{tot},\text{ss}}\delta F_{j}\right)e^{iH^{\text{el}}t}\right], \label{eq: diffusion calculation full 1}
\end{align}
with 
\begin{align}
\delta F_{i} & = \frac{\partial H^{\text{el}}}{\partial x_{i}} - F^{}_{i}(\mathbf{x}).
\end{align}
Here, $\text{Tr}_{\text{tot},\text{el}}(\dots)$ and $\rho^{\text{el}}_{\text{tot},\text{ss}}$ are the trace over and the steady state density matrix of all electronic degrees of freedom in the system and bath, respectively. This is in contrast to Eqs.\eqref{eq: average electronic force HEOM 1}-\eqref{eq: diffusion calculation HEOM} in the main text, where the trace is only over the system electronic degrees of freedom and $\boldsymbol{\tilde{\sigma}}^{\text{el}}_{\text{ss}}$ contains the steady state system electronic density matrix as well as all steady state ADOs. 

The first question, then, is whether the information one extracts from incorporating the bath effects via the ADOs is the same as that from treating the full electronic density matrix of system and bath. This is, however, not the only difference between the approaches. Eq.\eqref{eq: friction calculation full 1} and Eq.\eqref{eq: diffusion calculation full 1} contain the time-evolution of all electronic degrees of freedom  according to the full Hamiltonian, $H^{\text{el}}$, whereas Eq.\eqref{eq: friction calculation full 1} and Eq.\eqref{eq: diffusion calculation full 1} contain the time-evolution according to a HEOM derived under the assumption that at the initial state of the object being propagated, $A_{T}(0)$, can be factorized into system and bath components, $A_{T}(0) = A \otimes A_{B}$. Since the quantities $\frac{\partial \boldsymbol{\tilde{\sigma}}^{\text{el}}_{\text{ss}}}{\partial x_{j}}$ and $\delta F_{j}(\mathbf{x})\boldsymbol{\tilde{\sigma}}^{\text{el}}_{\text{ss}}$ evidently cannot be factorized, this raises another question; is the HEOM propagation defined by $\mathcal{L}^{\text{el}}(\mathbf{x})$ correct for these quantities? In the remainder of this section, both of these questions will be answered and the equivalence of the two forms of the current-induced forces will be shown.

\subsection{Average electronic force}\label{app: subsec: Average electronic force}

As in the main text, the key assumption is that all vibrational degrees of freedom are restricted to the system only, such that 
\begin{align}
\frac{\partial H^{\text{el}}(\mathbf{x})}{\partial x_{i}} & = \frac{\partial H^{\text{el}}_{\text{S}}(\mathbf{x})}{\partial x_{i}}
\end{align}
To start, consider the average electronic force, $F^{}_{i}(\mathbf{x})$. Because the spatial derivative now only involves system operators, it can be simplified to 
\begin{align}
F^{}_{i}(\mathbf{x}) & = -\text{Tr}_{\text{el}}\left[\frac{\partial H^{\text{el}}_{S}}{\partial x_{i}}\text{Tr}_{B}\left\{\rho^{\text{el}}_{\text{tot},\text{ss}}\right\}\right] = -\text{Tr}_{\text{el}}\left[\frac{\partial H^{\text{el}}_{S}}{\partial x_{i}}\sigma^{\text{el}}_{\text{ss}}\right], \label{eq: average force calculation full 1}
\end{align}
which is exactly the result one obtains from the HEOM approach in Eq.\eqref{eq: average electronic force HEOM 1}. The system electronic steady state density matrix obtained by HEOM is, by definition, the same as that obtained by the time-evolution of all electronic degrees of freedom to the steady state and then tracing out the bath, because HEOM is numerically exact and all systems considered have a unique steady state. 

\subsection{Friction tensor}\label{app: subsec: Friction tensor}

\begin{figure*}
\includegraphics[width = 0.9\textwidth]{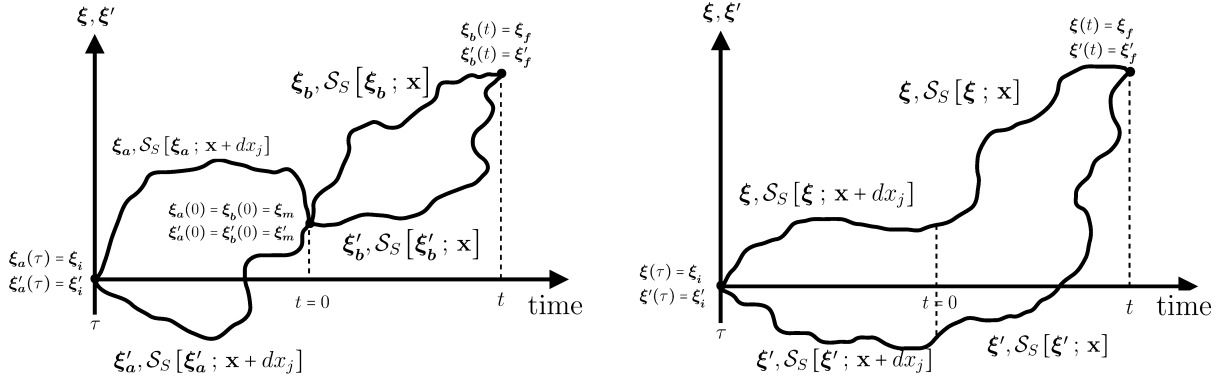}
\begin{centering}
\caption{Schematic of the transformation of the forward paths into one joint path. In the left plot, the paths $(\boldsymbol{\xi^{}_{a}},\boldsymbol{\xi^{\prime}_{a}})$ and $(\boldsymbol{\xi^{}_{b}},\boldsymbol{\xi^{\prime}_{b}})$ are joined at $t = 0$ by the midpoint $(\boldsymbol{\xi^{}_{m}},\boldsymbol{\xi^{\prime}_{m}})$. In Eq.\eqref{eq: A in coherent state representation}, however, this midpoint is integrated over and the two paths combine into one, albeit with different actions before and after $t = 0$, which is shown in the right plot.}
\label{fig: 7}
\end{centering}
\end{figure*}

Next, consider the more difficult quantities defined in Eq.\eqref{eq: friction calculation full 1} and Eq.\eqref{eq: diffusion calculation full 1}. In what follows, the equivalence will be shown explicitly for the friction, as the diffusion follows a very similar process. Similar to the previous subsection, Eq.\eqref{eq: friction calculation full 1} can be rewritten as 
\begin{align}
\gamma_{ij}(\mathbf{x}) & = -\int^{\infty}_{0} dt \: \text{Tr}_{\text{el}}\left[\frac{\partial H^{\text{el}}_{S}}{\partial x_{i}}A(\mathbf{x},t)\right], \label{eq: friction calculation full 2}
\end{align}
where
\begin{align}
A(\mathbf{x},t) & = \text{Tr}_{B}\left[e^{-iH^{\text{el}}(\mathbf{x})t}\frac{\partial \rho^{\text{el}}_{\text{tot},\text{ss}}}{\partial x_{j}}e^{iH^{\text{el}}(\mathbf{x})t}\right], \label{eq: quantity we need to calculate for equivalence}
\end{align}
has been defined; it is evidently this quantity that now must be shown to be equivalent to the HEOM approach. Note that the time-evolution has been explicitly written only as a function of the vibrational coordinates, $\mathbf{x}$, because the classical vibrational momenta do not couple directly to the electronic degrees of freedom. 

The next step, then, is to introduce the formal definition of the spatial derivative,
\begin{align}
\frac{\partial \rho^{\text{el}}_{\text{tot},\text{ss}}}{\partial x_{j}} & = \lim_{dx_{j} \rightarrow 0^{+}}\frac{\rho^{\text{el}}_{\text{tot},\text{ss}}(\mathbf{x}+dx_{j}) - \rho^{\text{el}}_{\text{tot},\text{ss}}(\mathbf{x)}}{dx_{j}}. \label{eq: rewrite spatial derivative}
\end{align}
Because the electronic systems under consideration have unique steady states, $\rho^{\text{el}}_{\text{tot},\text{ss}}(\mathbf{x}+dx_{j})$ and $\rho^{\text{el}}_{\text{tot},\text{ss}}(\mathbf{x})$ can be found by starting the total system at some point in the distant past, where they are assumed to be uncorrelated, $\lim\limits_{\tau \rightarrow -\infty}\rho^{\text{el}}_{\text{tot}}(\tau) = \rho^{\text{el}}(\tau)\rho^{\text{el}}_{B}(\tau)$, and evolving them up until time $t = 0$, such that 
\begin{align}
\rho^{\text{el}}_{\text{tot},\text{ss}}(\mathbf{x}+dx_{j}) & =  \lim_{\tau \rightarrow -\infty} e^{iH^{\text{el}}(\mathbf{x} + dx_{j})\tau}\rho^{\text{el}}(\tau)\rho^{\text{el}}_{B}(\tau)e^{-iH^{\text{el}}(\mathbf{x} + dx_{j})\tau} \\
\rho^{\text{el}}_{\text{tot},\text{ss}}(\mathbf{x}) & = \lim_{\tau \rightarrow -\infty} e^{iH(x)\tau}\rho^{\text{el}}(\tau)\rho^{\text{el}}_{B}(\tau)e^{-iH(x)\tau}.
\end{align}
Eq.\eqref{eq: quantity we need to calculate for equivalence}, therefore, will contain two terms, each with two sets of time-evolution. Consider the first term,
\begin{align}
A_{\mathbf{x} + dx_{j}}(\mathbf{x},t) & = \text{Tr}_{B}\left[e^{-iH^{\text{el}}(\mathbf{x})t}\rho^{\text{el}}_{\text{tot},\text{ss}}(\mathbf{x}+dx_{j})e^{iH^{\text{el}}(\mathbf{x})t}\right] \\
& = \lim_{\tau \rightarrow -\infty} \text{Tr}_{B}\left[e^{-iH^{\text{el}}(\mathbf{x})t}e^{iH^{\text{el}}(\mathbf{x} + dx_{j})\tau} \right. \nonumber \\
& \hspace{1.5cm} \left.\rho^{\text{el}}(\tau)\rho^{\text{el}}_{B}(\tau)e^{-iH^{\text{el}}(\mathbf{x} + dx_{j})\tau}e^{iH^{\text{el}}(\mathbf{x})t}\right],
\end{align}
which can now be evaluated using HEOM. Note that the subscript notation containing $\mathbf{x} + dx_{j}$ specifies the vibrational coordinates as the electronic degrees of freedom evolve to the steady state from $\tau$ to $t = 0$. It is only this part that needs to be specified, because, from Eq.\eqref{eq: friction calculation full 2}, one can see that the evolution after $t = 0$ for any quantity is always at $\mathbf{x}$. 

Not all steps of the standard HEOM derivation will be reproduced here, as they can be found in Refs. \cite{Tanimura1989,Tanimura2006,Haertle2015,Jin2007,Jin2008,Xiao2009,Yan2014,Wenderoth2016,Tanimura2020,Ye2016}. First, the basis of fermionic coherent states, which are eigenvectors of the fermionic annihilation and creation operators, is introduced
\begin{align}
d^{}_{i}|\xi\rangle & = \xi_{i}|\xi\rangle \hspace{0.5cm} \text{ and } \hspace{0.5cm} \langle \xi |d^{\dag}_{i} = \langle \xi |\xi_{i}^{*}.
\end{align}
The eigenvalues $\xi_{m}$, which are often collected into vectors, $\boldsymbol{\xi} = \left(\xi_{1},\xi_{2},\dots{}\right)$ and $\boldsymbol{\xi^{*}} = \left(\xi^{*}_{1},\xi^{*}_{2},\dots{}\right)$, are not ordinary complex numbers, but rather Grassmann variables satisfying the anti-commutation relation between fermionic annihilation and creation operators. In this basis, $A_{\mathbf{x} + dx_{j}}(\mathbf{x},t)$ is 
\begin{widetext}
\begin{align}
A_{\mathbf{x} + dx_{j}}(\boldsymbol{\xi^{}_{f}},\boldsymbol{\xi^{\prime}_{f}} ; \mathbf{x},t) & = \lim_{\tau \rightarrow -\infty} \int (d \boldsymbol{\xi_{i}^{*}}d \boldsymbol{\xi_{i}^{}})\:(d \boldsymbol{\xi_{i}^{*\prime}}d \boldsymbol{\xi_{i}^{\prime}}) \:(d \boldsymbol{\xi_{i_{B}}^{*}}d \boldsymbol{\xi^{}_{i_{B}}})\:(d \boldsymbol{\xi_{i_{B}}^{*\prime}}d \boldsymbol{\xi^{\prime}_{i_{B}}})e^{-\boldsymbol{\xi_{i}^{*}}\boldsymbol{\xi^{}_{i}}}e^{-\boldsymbol{\xi_{i}^{*\prime}}\boldsymbol{\xi^{\prime}_{i}}} e^{-\boldsymbol{\xi_{i_{B}}^{*}}\boldsymbol{\xi^{}_{i_{B}}}}e^{-\boldsymbol{\xi_{i_{B}}^{*\prime}}\boldsymbol{\xi^{\prime}_{i_{B}}}}  \nonumber \\
& \hspace{1.3cm} \int (d \boldsymbol{\xi_{m}^{*}}d \boldsymbol{\xi_{m}^{}})\:(d \boldsymbol{\xi_{m}^{*\prime}}d \boldsymbol{\xi^{\prime}_{m}})(d \boldsymbol{\xi_{m_{B}}^{*}}d \boldsymbol{\xi_{m_{B}}})\:(d \boldsymbol{{\xi_{m_{B}}^{*\prime}}}d \boldsymbol{\xi^{\prime}_{m_{B}}}) e^{-\boldsymbol{\xi_{m}^{*}}\boldsymbol{\xi^{}_{m}}}e^{-\boldsymbol{\xi_{m}^{*\prime}}\boldsymbol{\xi^{\prime}_{m}}}  e^{-\boldsymbol{\xi_{m_{B}}^{*}}\boldsymbol{\xi^{}_{m_{B}}}}e^{-\boldsymbol{\xi_{m_{B}}^{*\prime}}\boldsymbol{\xi^{\prime}_{m_{B}}}} \nonumber \\
& \hspace{1.3cm} \text{Tr}_{B}\Big[G_{\mathbf{x}}(\boldsymbol{\xi_{f_{B}}},\boldsymbol{\xi^{}_{f}},t ; \boldsymbol{\xi^{}_{m_{B}}},\boldsymbol{\xi^{}_{m}}, 0)G_{\mathbf{x} + dx_{j}}(\boldsymbol{\xi_{m_{B}}},\boldsymbol{\xi_{m}}, 0; \boldsymbol{\xi_{i_{B}}},\boldsymbol{\xi_{i}}, \tau) \nonumber \\
& \hspace{2cm} \langle \xi_{i_{B}},\xi_{i}|\rho^{\text{el}}(\tau)\rho^{\text{el}}_{B}(\tau)|\xi^{\prime}_{i},\xi^{\prime}_{i_{B}}\rangle G^{*}_{\mathbf{x} + dx_{j}}(\boldsymbol{\xi^{\prime}_{m_{B}}},\boldsymbol{\xi_{m}^{\prime}}, 0; \boldsymbol{\xi^{\prime}_{i_{B}}},\boldsymbol{\xi^{\prime}_{i}} , \tau ) G^{*}_{\mathbf{x}}(\boldsymbol{\xi^{\prime}_{f_{B}}},\boldsymbol{\xi^{\prime}_{f}},t ; \boldsymbol{\xi_{m_{B}}^{\prime}},\boldsymbol{\xi^{\prime}_{m}} ,0 )\Big]. \label{eq: A in coherent state representation}
\end{align}
\end{widetext}
In Eq.\eqref{eq: A in coherent state representation}, coherent states with a $B$ subscript denote bath variables, while those without a subscript denote system variables. The object, 
\begin{align}
G_{\mathbf{x}}(\boldsymbol{\xi_{f_{B}}},\boldsymbol{\xi^{}_{f}},t ; \boldsymbol{\xi^{}_{m_{B}}},\boldsymbol{\xi^{}_{m}}, 0) & = \langle \xi_{f_{B}},\xi^{}_{f} | e^{-iH^{\text{el}}(\mathbf{x})t} | \xi^{}_{m},\xi^{}_{m_{B}} \rangle,
\end{align}
for example, denotes forward time propagation from coherent states $|\xi_{m_{B}},\xi_{m}\rangle$ at time $t = 0$ to coherent states $|\xi_{f_{B}},\xi_{f}\rangle$ at time $t$, all while the system is held at vibrational position $\mathbf{x}$. 

Since the total density matrix factorizes in the infinite past, $\rho^{\text{el}}(\tau)$ can be brought outside the bath trace, and the expression can be written with a propagator,
\begin{widetext}
\begin{align}
A_{\mathbf{x} + dx_{j}}(\boldsymbol{\xi^{}_{f}},\boldsymbol{\xi^{\prime}_{f}} ; \mathbf{x}, t) &  =  \lim_{\tau \rightarrow -\infty} \int (d \boldsymbol{\xi_{i}^{*}}d \boldsymbol{\xi_{i}^{}})\:(d \boldsymbol{\xi_{i}^{*\prime}}d \boldsymbol{\xi_{i}^{\prime}})(d \boldsymbol{\xi_{m}^{*}}d \boldsymbol{\xi_{m}^{*}})\:(d \boldsymbol{\xi_{m}^{*\prime}}d \boldsymbol{\xi_{m}^{\prime}}) e^{-\boldsymbol{\xi_{i}^{*}}\boldsymbol{\xi_{i}}}e^{-\boldsymbol{\xi_{i}^{*\prime}}\boldsymbol{\xi^{\prime}_{i}}}e^{-\boldsymbol{\xi_{m}^{*}}\boldsymbol{\xi^{}_{m}}}e^{-\boldsymbol{\xi_{m}^{*\prime}}\boldsymbol{\xi_{m}^{\prime}}} \nonumber \\
& \hspace{1.5cm} \times\mathcal{J}_{\mathbf{x}+dx_{j}}\left(\boldsymbol{\xi^{}_{f}},\boldsymbol{\xi^{\prime}_{f}}, t ; \boldsymbol{\xi_{m}},\boldsymbol{\xi^{\prime}_{m}}, 0 ; \boldsymbol{\xi_{i}},\boldsymbol{\xi^{\prime}_{i}}, \tau \right) \rho^{\text{el}}(\boldsymbol{\xi}_{i},\boldsymbol{\xi}'_{i} ; \tau). \label{eq: propagator 1}
\end{align}
\end{widetext}
The propagator is then decomposed into four path integrals over the system coherent states, with path $(\boldsymbol{\xi^{}_{a}},\boldsymbol{\xi^{*}_{a}})$ from $\tau$ to the steady state at time $t = 0$, and path $(\boldsymbol{\xi^{}_{b}},\boldsymbol{\xi^{*}_{b}})$ from $t = 0$ to time $t$,
\begin{widetext}
\begin{align}
 \mathcal{J}_{\mathbf{x}+dx_{j}}\left(\boldsymbol{\xi^{}_{f}},\boldsymbol{\xi^{\prime}_{f}}, t ; \boldsymbol{\xi_{m}},\boldsymbol{\xi^{\prime}_{m}}, 0 ; \boldsymbol{\xi_{i}},\boldsymbol{\xi^{\prime}_{i}}, \tau \right) & = \int_{\boldsymbol{\xi_{b}}(0) = \boldsymbol{\xi_{m}}}^{\boldsymbol{\xi_{b}^{*}}(t) = \boldsymbol{\xi^{*}_{f}}}\mathcal{D}\left(\boldsymbol{\xi^{}_{b}},\boldsymbol{\xi_{b}^{*}}\right)\int_{\boldsymbol{\xi^{\prime}_{b}}(0) = \boldsymbol{\xi^{\prime}_{m}}}^{\boldsymbol{\xi^{*\prime}_{b}}(t) = \boldsymbol{\xi^{*\prime}_{f}}} \mathcal{D}\left(\boldsymbol{\xi_{b}^{\prime}},{\boldsymbol{\xi^{*\prime}_{b}}}\right) \int_{\boldsymbol{\xi^{}_{a}}(\tau) = \boldsymbol{\xi^{}_{i}}}^{\boldsymbol{\xi_{a}^{*}}(0) = \boldsymbol{\xi^{*}_{m}}}\mathcal{D}\left(\boldsymbol{\xi^{}_{a}},\boldsymbol{\xi_{a}^{*}}\right)\int_{\boldsymbol{\xi^{\prime}_{a}}(\tau) = \boldsymbol{\xi^{\prime}_{i}}}^{\boldsymbol{\xi^{*\prime}_{a}}(0) = \boldsymbol{\xi^{*\prime}_{m}}} \mathcal{D}\left(\boldsymbol{\xi_{a}}',{\boldsymbol{\xi^{*\prime}_{a}}}\right) \nonumber \\ 
& \:\:\:\:\: \times e^{i\mathcal{S}_{S}\left[\boldsymbol{\xi^{*}_{b}},\boldsymbol{\xi^{}_{b}} \: ; \: \mathbf{x},t,0 \right]}e^{i\mathcal{S}_{S}\left[\boldsymbol{\xi^{*}_{a}},\boldsymbol{\xi^{}_{a}} \: ; \: \mathbf{x} + dx_{j},0,\tau\right]}  \mathcal{F}[\boldsymbol{\xi^{}_{a}},\boldsymbol{\xi^{\prime}_{a}},\boldsymbol{\xi^{*}_{a}},\boldsymbol{\xi^{*\prime}_{a}},\boldsymbol{\xi^{}_{b}},\boldsymbol{\xi^{\prime}_{b}},\boldsymbol{\xi^{*}_{b}},\boldsymbol{\xi^{*\prime}_{b}} ; t,\tau] \nonumber \\ 
& \:\:\:\:\: \times e^{-i\mathcal{S}^{*}_{S}\left[\boldsymbol{\xi^{*\prime}_{a}},\boldsymbol{\xi^{\prime}_{a}} \: ; \: \mathbf{x} + dx_{j} , 0 , \tau \right]}e^{-i\mathcal{S}^{*}_{S}\left[\boldsymbol{\xi^{*\prime}_{b}},\boldsymbol{\xi^{\prime}_{b}} \: ; \: \mathbf{x},t,0\right]}.
\end{align}
\end{widetext}
A schematic of these four path integrals for the forward propagation is shown in the left plot of Fig. \ref{fig: 7}, where paths $(\boldsymbol{\xi^{}_{a}},\boldsymbol{\xi^{*}_{a}})$ and $(\boldsymbol{\xi^{}_{b}},\boldsymbol{\xi^{*}_{b}})$ are joined by the midle point, $(\boldsymbol{\xi^{}_{m}},\boldsymbol{\xi^{*}_{m}})$. The effect of the bath on each path is contained in the Feynman-Vernon influence functional, $\mathcal{F}[\boldsymbol{\xi^{}_{a}},\boldsymbol{\xi^{\prime}_{a}},\boldsymbol{\xi^{*}_{a}},\boldsymbol{\xi^{*\prime}_{a}},\boldsymbol{\xi^{}_{b}},\boldsymbol{\xi^{\prime}_{b}},\boldsymbol{\xi^{*}_{b}},\boldsymbol{\xi^{*\prime}_{b}} ; t,\tau]$. Notice that, because the vibrational coordinates enter only through $H^{\text{el}}_{S}$, the influence-functional is the same for path $(\boldsymbol{\xi^{}_{a}},\boldsymbol{\xi^{*}_{a}})$ and path $(\boldsymbol{\xi^{}_{b}},\boldsymbol{\xi^{*}_{b}})$. The difference between the two sets of paths, then, resides exclusively in the system actions,
\begin{widetext}
\begin{align}
\mathcal{S}_{S}\left[\boldsymbol{\xi^{*}_{b}},\boldsymbol{\xi^{}_{b}} \: ; \: \mathbf{x}, t,0\right] & = -i \boldsymbol{\xi^{*}_{b}}(t),\boldsymbol{\xi^{}_{b}}(t) + \int^{t}_{0} d\tau_{1} \left[i\boldsymbol{\xi^{*}_{b}}(\tau_{1})\frac{\partial \boldsymbol{\xi^{}_{b}}(\tau_{1})}{\partial \tau_{1}} - H^{\text{el}}_{S}\left(\boldsymbol{\xi^{}_{b}}(\tau_{1}),\boldsymbol{\xi^{*}_{b}}(\tau_{1}) \: ; \: \mathbf{x}\right)\right] \\
\mathcal{S}_{S}\left[\boldsymbol{\xi^{*}_{a}},\boldsymbol{\xi^{}_{a}} \: ; \: \mathbf{x} + dx_{j} , 0,\tau\right] & = -i \boldsymbol{\xi^{*}_{a}}(0),\boldsymbol{\xi^{}_{a}}(0) + \int^{0}_{\tau} d\tau_{1} \left[i\boldsymbol{\xi^{*}_{a}}(\tau_{1})\frac{\partial \boldsymbol{\xi^{}_{a}}(\tau_{1})}{\partial \tau_{1}}  - H^{\text{el}}_{S}\left(\boldsymbol{\xi^{}_{a}}(\tau_{1}),\boldsymbol{\xi^{*}_{a}}(\tau_{1})\: ; \: \mathbf{x} + dx_{j}\right)\right].
\end{align}
\end{widetext}
This greatly simplifies the theory, as now the two distinct paths can be joined by the integral over the middle point in Eq.\eqref{eq: A in coherent state representation}. The resulting joint path is shown in the right plot in Fig. \ref{fig: 7}, with the total propagator and influence functional now 
\begin{widetext}
\begin{align}
\mathcal{J}_{\mathbf{x}+dx_{j}}\left(\boldsymbol{\xi^{}_{f}},\boldsymbol{\xi}'_{f},t ; \boldsymbol{\xi}_{i},\boldsymbol{\xi}'_{i},\tau \right) & = \int_{\boldsymbol{\xi}(\tau) = \boldsymbol{\xi_{i}}}^{\boldsymbol{\xi^{*}}(t) = \boldsymbol{\xi^{*}_{f}}}\mathcal{D}\left(\boldsymbol{\xi},\boldsymbol{\xi^{*}}\right)\int_{\boldsymbol{\xi^{\prime}}(\tau) = \boldsymbol{\xi^{\prime}_{i}}}^{\boldsymbol{\xi^{*\prime}}(t) = \boldsymbol{\xi^{*\prime}_{f}}} \mathcal{D}\left(\boldsymbol{\xi^{\prime}},{\boldsymbol{\xi^{*\prime}}}\right) \nonumber \\
& \:\:\:\:\: e^{i\mathcal{S}_{S}\left[\boldsymbol{\xi^{*}},\boldsymbol{\xi} \: ; \: \mathbf{x},t , 0\right]}e^{i\mathcal{S}_{S}\left[\boldsymbol{\xi^{*}},\boldsymbol{\xi} \: ; \: \mathbf{x} + dx_{j},0,\tau\right]}\mathcal{F}[\boldsymbol{\xi^{}},\boldsymbol{\xi^{\prime}},\boldsymbol{\xi^{*}},\boldsymbol{\xi^{*\prime}} ; t,\tau]  e^{-i\mathcal{S}^{*}_{S}\left[\boldsymbol{\xi^{*\prime}},\boldsymbol{\xi^{\prime}} \: ; \: \mathbf{x} + dx_{j} , 0, \tau \right]}e^{-i\mathcal{S}^{*}_{S}\left[\boldsymbol{\xi^{*\prime}},\boldsymbol{\xi^{\prime}} \: ; \: \mathbf{x},t,0\right]} \\
\mathcal{F}[\boldsymbol{\xi^{}},\boldsymbol{\xi^{\prime}},\boldsymbol{\xi^{*}},\boldsymbol{\xi^{*\prime}};t,\tau] & = \int (d \boldsymbol{\xi_{i_{B}}^{*}}d \boldsymbol{\xi^{}_{i_{B}}})\:(d \boldsymbol{\xi_{i_{B}}^{*\prime}}d \boldsymbol{\xi_{i_{B}}^{\prime}}) \: e^{-\boldsymbol{\xi_{i_{B}}^{*}}\boldsymbol{\xi^{}_{i_{B}}}}e^{-\boldsymbol{\xi_{i_{B}}^{*\prime}}\boldsymbol{\xi^{\prime}_{i_{B}}}} \nonumber \\
& \:\:\:\:\: \text{Tr}_{B}\Big[G_{B}(\boldsymbol{\xi^{}_{f_{B}}},\boldsymbol{\xi^{}_{f}},t ; \boldsymbol{\xi^{}_{i_{B}}},\boldsymbol{\xi^{}_{i}} , \tau )\langle \xi^{}_{i_{B}}|\rho_{B}(0)|\xi^{\prime}_{i_{B}}\rangle  G_{B}^{*}(\boldsymbol{\xi^{\prime}_{f_{B}}},\boldsymbol{\xi^{\prime}_{f}},t ; \boldsymbol{\xi^{\prime}_{i_{B}}},\boldsymbol{\xi^{\prime}_{i}} , \tau )\Big].
\end{align} 
\end{widetext}
Here, $G_{B}(\boldsymbol{\xi^{}_{f_{B}}},\boldsymbol{\xi^{}_{f}},t ; \boldsymbol{\xi^{}_{i_{B}}},\boldsymbol{\xi^{}_{i}} ,0 )$ is the propagator for the bath density matrix while the system evolves along path $(\boldsymbol{\xi^{*}},\boldsymbol{\xi})$. 

In the bath interaction picture, where $H_{SB}^{I}(t) = e^{-iH_{B}t}H_{SB}e^{iH_{B}t}$, this time-evolution can be alternatively expressed with time-ordered exponentials,
\begin{align}
\mathcal{F} & = \text{Tr}_{B} \left\{ \mathcal{T} e^{-i\int^{t}_{\tau} d\tau_{1} H_{SB}^{I}(\tau_{1} \: ; \: \boldsymbol{\xi^{}}(\tau_{1}),\boldsymbol{\xi^{*}}(\tau_{1}))}\rho^{\text{el}}_{B}(\tau) \right.\nonumber \\
& \hspace{1cm} \left.\times \mathcal{T}^{-1} e^{i\int^{t}_{\tau} d\tau_{1} H_{SB}^{I}(\tau_{1} \: ; \: \boldsymbol{\xi^{\prime}}(\tau_{1}),{\boldsymbol{\xi^{*\prime}}}(\tau_{1}))}\right\}, \label{eq: influence-functional 1}
\end{align}
where the coherent state notation has been suppressed from $\mathcal{F}$ for brevity. 

The influence functional in Eq.\eqref{eq: influence-functional 1} is similar to that found in standard HEOM theory, except now the initial time is at some point in the distant past, $\tau$. Following the usual HEOM approach \cite{Tanimura1989,Tanimura2006,Haertle2015,Jin2007,Jin2008,Xiao2009,Yan2014,Wenderoth2016,Tanimura2020,Ye2016} allows the influence functional to be expressed in terms of the two-time bath-correlation functions, $C^{\sigma}_{\alpha,mm'}(t)$, and their expansion from Eq.\eqref{eq: exponential decomposition} in the main text,
\begin{align}
\mathcal{F} & = \exp\Big(-i \int^{t}_{\tau} d\tau_{1} \sum_{j} \mathcal{A}^{\bar{\sigma}}_{\alpha,m}\left[\tau_{1} ; \boldsymbol{\xi},\boldsymbol{\xi}'\right] \mathcal{B}_{j}\left[\tau_{1},\tau ; \boldsymbol{\xi},\boldsymbol{\xi}'\right] \Big).
\end{align}
Here, two new Grassmann variables have been introduced,
\begin{align}
\mathcal{B}_{j}\left[\boldsymbol{\xi},\boldsymbol{\xi}'  ; t,\tau\right] & = -iV^{}_{\alpha,m}\left(\eta^{}_{j} \int^{t}_{\tau}d\tau_{1} \: e^{-\kappa^{}_{j}\tau_{1}}{\xi}^{\sigma}_{m}(t - \tau_{1}) \right. \nonumber \\
& \hspace{2cm} \left. - \eta^{*}_{j} \int^{t}_{\tau}d\tau_{1} \: e^{-\kappa^{}_{j}\tau_{1}}{\xi}^{\sigma\prime}_{m}(t - \tau_{1})\right) \\
 \mathcal{A}^{\sigma}_{\alpha,m}\left[\boldsymbol{\xi},\boldsymbol{\xi}'  ; t \right] & = V^{}_{\alpha,m} \left[\xi^{\sigma}_{m}(t) + \xi^{\sigma\prime}_{m}(t)\right],
\end{align}
along with the super-index $j = (\alpha,\sigma,\ell,m)$ from the main text. One can now construct the standard hierarchical equations of motion for $\mathcal{F}$,
\begin{align}
\frac{\partial}{\partial t}\mathcal{F}^{(n)}_{\mathbf{j}} & =  - \left(\sum_{r = 1}^{N} \kappa^{}_{j_{r}}\right)\mathcal{F}^{(n)}_{\mathbf{j}} - i\sum_{r=1}^{n} (-1)^{n-r}\mathcal{C}_{j_{r}}\mathcal{F}^{(n-1)}_{\mathbf{j}^{-}} \nonumber \\
& \hspace{0.5cm} - i\sum_{j} \mathcal{A}^{\bar{\sigma}}_{\alpha,m} \mathcal{F}^{(n+1)}_{\mathbf{j}^{+}}, \label{eq: HEOM Final F}
\end{align}
where the $n$th-tier influence functional is formed by application of $n$ Grassmann variables $\mathcal{B}_{j}\left[\boldsymbol{\xi^{}},\boldsymbol{\xi^{\prime}} ; t,\tau\right]$,
\begin{align}
\mathcal{F}^{(n)}_{\mathbf{j}} & = \mathcal{B}_{j_{n}}\dots\mathcal{B}_{j_{1}}\mathcal{F},
\end{align}
such that the $0$th tier corresponds to the original influence functional, $\mathcal{F}^{(0)}  = \mathcal{F}$.

Now explicitly including the coherent state notation, a corresponding $n$th tier propagator and auxiliary operator (AO) can be respecetively defined,
\begin{widetext}
\begin{align}
\mathcal{J}^{(n)}_{\mathbf{j},\mathbf{x}+dx_{j}}\left(\boldsymbol{\xi^{}_{f}},\boldsymbol{\xi}'_{f},t ; \boldsymbol{\xi}_{i},\boldsymbol{\xi}'_{i},\tau \right) & = \int_{\boldsymbol{\xi}(\tau) = \boldsymbol{\xi_{i}}}^{\boldsymbol{\xi^{*}}(t) = \boldsymbol{\xi^{*}_{f}}}\mathcal{D}\left(\boldsymbol{\xi},\boldsymbol{\xi^{*}}\right)\int_{\boldsymbol{\xi^{\prime}}(\tau) = \boldsymbol{\xi^{\prime}_{i}}}^{\boldsymbol{\xi^{*\prime}}(t) = \boldsymbol{\xi^{*\prime}_{f}}} \mathcal{D}\left(\boldsymbol{\xi^{\prime}},{\boldsymbol{\xi^{*\prime}}}\right) e^{i\mathcal{S}_{S}\left[\boldsymbol{\xi^{*}},\boldsymbol{\xi} \: ; \: \mathbf{x},t,0 \right]}e^{i\mathcal{S}_{S}\left[\boldsymbol{\xi^{*}},\boldsymbol{\xi} \: ; \: \mathbf{x} + dx_{j},0,\tau\right]} \nonumber \\
& \:\:\:\:\: \times\mathcal{F}^{(n)}_{\mathbf{j}}[\boldsymbol{\xi^{}},\boldsymbol{\xi^{\prime}},\boldsymbol{\xi^{*}},\boldsymbol{\xi^{*\prime}} ; t,\tau]e^{-i\mathcal{S}^{*}_{S}\left[\boldsymbol{\xi^{*\prime}},\boldsymbol{\xi^{\prime}} \: ; \: \mathbf{x} + dx_{j} , 0 ,\tau\right]}e^{-i\mathcal{S}^{*}_{S}\left[\boldsymbol{\xi^{*\prime}},\boldsymbol{\xi^{\prime}} \: ; \: \mathbf{x},t,0\right]} \\
A^{(n)}_{\mathbf{j},\mathbf{x} + dx_{j}}(\boldsymbol{\xi^{}_{f}},\boldsymbol{\xi^{\prime}_{f}} ; \mathbf{x}, t) & = \lim_{\tau \rightarrow -\infty} \int (d \boldsymbol{\xi_{i}^{*}}d \boldsymbol{\xi_{i}^{}})\:(d \boldsymbol{\xi_{i}^{*\prime}}d \boldsymbol{\xi_{i}^{\prime}}) e^{-\boldsymbol{\xi_{i}^{*}}\boldsymbol{\xi_{i}}}e^{-\boldsymbol{\xi_{i}^{*\prime}}\boldsymbol{\xi^{\prime}_{i}}}\mathcal{J}^{(n)}_{\mathbf{j},\mathbf{x}+dx_{j}}\left(\boldsymbol{\xi^{}_{f}},\boldsymbol{\xi^{\prime}_{f}},t ; \boldsymbol{\xi^{}_{i}},\boldsymbol{\xi^{\prime}_{i}},\tau \right) \rho^{\text{el}}(\boldsymbol{\xi^{}_{i}},\boldsymbol{\xi^{\prime}_{i}} ; \tau). \label{eq: propagator 2}
\end{align}
\end{widetext}

Consider now the case at $t = 0$. In this instance, two of the four system actions vanish, $\mathcal{S}_{S}\left[\boldsymbol{\xi^{*}},\boldsymbol{\xi} \: ; \: \mathbf{x}, 0,0\right] = 0$, and, after moving to a more computationally suitable basis, the AOs propagate according to 
\begin{widetext}
\begin{align}
\frac{\partial}{\partial t'}{A}^{(n)}_{\mathbf{j},\mathbf{x} + dx_{j}}(t') &  = i\left[H_{\text{S}}(\mathbf{x} + dx_{j}),A^{(n)}_{\mathbf{j},\mathbf{x} + dx_{j}}\right] - \left(\sum_{r = 1}^{N} \kappa^{}_{j_{r}}\right)A^{(n)}_{\mathbf{j},\mathbf{x} + dx_{j}} + i\sum_{r=1}^{n} (-1)^{n-r}\mathcal{C}_{j_{r}}A^{(n-1)}_{\mathbf{j}^{-},\mathbf{x} + dx_{j}} + i\sum_{j} \mathcal{A}^{\bar{\sigma}}_{\alpha,m} A^{(n+1)}_{\mathbf{j}^{+},\mathbf{x} + dx_{j}}, \label{eq: HEOM useful 1}
\end{align}
\end{widetext}
where $\tau < t' \leq 0$. Eq.\eqref{eq: HEOM useful 1} is solved subject to the initial condition $A^{(0)}_{\mathbf{x} + dx_{j}}(\tau) = \rho^{\text{el}}(\tau)$, which is the initial state of the system electronic degrees of freedom in the distant past, and $A^{(n > 0)}_{\mathbf{j},\mathbf{x} + dx_{j}}(\tau) = 0$. If one solves these equations of motion for 
\begin{align}
\boldsymbol A(\mathbf{x} + dx_{j},0) & = \left[A^{(0)}_{\mathbf{x} + dx_{j}}(0),A^{(1)}_{j_{1},\mathbf{x} + dx_{j}}(0),\dots\right],
\end{align} 
then one is simply calculating 
\begin{align}
\boldsymbol{\tilde{\sigma}}^{\text{el}}_{\text{ss}}(\mathbf{x} + dx_{j}) & = \left[\sigma^{(0),\text{el}}_{\text{ss}}(\mathbf{x} + dx_{j}),\sigma^{(1),\text{el}}_{j_{1},\text{ss}}(\mathbf{x} + dx_{j}),\dots\right],
\end{align}
the steady of all ADOs for vibrational frame $\mathbf{x} + dx_{j}$. This entire process can be repeated for the second term in Eq.\eqref{eq: rewrite spatial derivative}, ${A}_{\mathbf{x}}(\mathbf{x},t)$, which just yields $\boldsymbol{\tilde{\sigma}}^{\text{el}}_{ss}(\mathbf{x})$. 

With these results, the propagation from time $t = 0$ to time $t$ can be easily treated. First, note that both ${A}_{\mathbf{x}}(\mathbf{x},t)$ and ${A}_{\mathbf{x} + dx_{j}}(\mathbf{x},t)$ will follow the same equation of motion for vibrational frame $\mathbf{x}$ during this interval. Since the AOs at $t = 0$ are known from the previous section, furthermore,
\begin{align}
\lim_{\tau \rightarrow -\infty} A^{(n)}_{\mathbf{j},\mathbf{x} + dx_{j}}(0,\tau) & = \sigma^{(n),\text{el}}_{\mathbf{j},\text{ss}}(\mathbf{x} + dx_{j}) \nonumber \\
\lim_{\tau \rightarrow -\infty} A^{(n)}_{\mathbf{j},\mathbf{x}}(0,\tau) & = \sigma^{(n),\text{el}}_{\mathbf{j},\text{ss}}(\mathbf{x}),
\end{align}
both can just be propagated from this point to time $t$. After putting both the $\mathbf{A}_{\mathbf{x} + dx_{j}}(\mathbf{x},t)$ and $\mathbf{A}_{\mathbf{x}}(\mathbf{x},t)$ back together, then, Eq.\eqref{eq: quantity we need to calculate for equivalence} becomes
\begin{widetext}
\begin{align}
A^{(n)}_{\mathbf{j},\mathbf{x} + dx_{j}}(\boldsymbol{\xi^{}_{f}},\boldsymbol{\xi^{\prime}_{f}} ; \mathbf{x}, t)&  =\lim_{\tau \rightarrow -\infty} \int (d \boldsymbol{\xi_{i}^{*}}d \boldsymbol{\xi_{i}^{}})\:(d \boldsymbol{\xi_{i}^{*\prime}}d \boldsymbol{\xi_{i}^{\prime}}) e^{-\boldsymbol{\xi_{i}^{*}}\boldsymbol{\xi_{i}}}e^{-\boldsymbol{\xi_{i}^{*\prime}}\boldsymbol{\xi^{\prime}_{i}}}  \mathcal{J}^{(n)}_{\mathbf{j},\mathbf{x}}\left(\boldsymbol{\xi^{}_{f}},\boldsymbol{\xi^{\prime}_{f}},t ; \boldsymbol{\xi^{}_{i}},\boldsymbol{\xi^{\prime}_{i}}, 0 \right) \frac{\partial \sigma^{(n),\text{el}}_{\mathbf{j},\text{ss}}(\boldsymbol{\xi^{}_{i}},\boldsymbol{\xi^{\prime}_{i}} ; \mathbf{x})}{\partial x_{j}}. 
\end{align}
\end{widetext}
Again, returning to a more computationally suitable basis yields the corresponding HEOM, 
\begin{align}
\frac{\partial}{\partial t}A^{(n)}_{\mathbf{j}}(\mathbf{x},t) & = -i\left[H_{\text{S}}(\mathbf{x}),A^{(n)}_{\mathbf{j}}\right] - \left(\sum_{r = 1}^{N} \kappa^{}_{j_{r}}\right)A^{(n)}_{\mathbf{j}} \nonumber \\
& \:\:\:\:\: -  i\sum_{r=1}^{n} (-1)^{n-r}\mathcal{C}_{j_{r}}A^{(n-1)}_{\mathbf{j}^{-}} \nonumber \\
& \:\:\:\:\: - i\sum_{j} \mathcal{A}^{\bar{\sigma}}_{\alpha,m} A^{(n+1)}_{\mathbf{j}^{+}}, \label{eq: HEOM useful 3}
\end{align}
an equation of motion that is solved subject to the initial condition 
\begin{align}
A^{(n)}_{\mathbf{j}}(\mathbf{x},0) & = \frac{\partial \boldsymbol{\tilde{\sigma}}^{\text{el}}_{ss}(\mathbf{x})}{\partial x_{j}}.
\end{align}
Note that this is exactly the same HEOM one would obtain for the time-evolution of a reduced system density matrix based on a factorized initial condition, except that now it time-evolves the spatial derivative of the steady state of all ADOs. 

Evidently, therefore, both questions from the start of this section have been answered. First, when the vibrational degrees of freedom are only in the system Hamiltonian, $H^{\text{el}}_{S}$, one extracts the same information from $\boldsymbol{\tilde{\sigma}}^{\text{el}}_{ss}(\mathbf{x})$ as from $\rho^{\text{el}}_{\text{tot},\text{ss}}(\mathbf{x})$. Second, the resulting time-propagation is under the same HEOM as that derived for a factorized initial state. Finally, to fully connect this with the form of the friction of the main text, consider the HEOM in Eq.\eqref{eq: HEOM useful 3} in a joint Liouville space, where it has the solution 
\begin{align}
\boldsymbol A(\mathbf{x},t) & = e^{-\mathcal{L}^{\text{el}}(\mathbf{x})t}\frac{\partial \boldsymbol{\tilde{\sigma}}^{\text{el}}_{ss}(\mathbf{x})}{\partial x_{j}}.
\end{align}
Here, $\mathcal{L}^{\text{el}}(\mathbf{x})$ is a superoperator containing the numerically exact dynamics in Eq.\eqref{eq: HEOM useful 3} and is, by definition, the same quantity as in Eq.\eqref{eq: QCLE final}. The friction in Eq.\eqref{eq: friction calculation full 1}, consequently, is now exactly the same expression as that derived in the main text. 

\subsection{Diffusion tensor}\label{app: subsec: Diffusion tensor}

The equivalence of the diffusion tensors can be shown through a similar process. The only difference is that an operator in the system space, $\delta F_{j}(\mathbf{x})$, is applied at time $t = 0$, instead of a classical spatial derivative, which is the same problem that the authors already considered in Ref. \cite{Ke2022}.

\section{Non-Markovian friction tensor}\label{app: Non-Markovian friction tensor}

The friction and diffusion tensors derived in the previous section are Markovian, in that they depend only on the vibrational position at time $t$. In general, introducing non-Markovian dynamics for this problem is nontrivial; however, a simple form of memory can be introduced by setting $\frac{\partial \mathbf{\tilde{B}}^{\text{el}}}{\partial t} \neq 0$, as in Ref. \cite{Dou2017c}. This still keeps the timescale separation between electronic and vibrational degrees of freedom, in that the time-evolution of $\mathbf{\tilde{B}}^{\text{el}}(\mathbf{x},\mathbf{p} ; t)$ and $A(\mathbf{x},\mathbf{p} ; t)$ is still for a fixed vibrational frame, but now some relaxation effects from the electronic degrees of freedom are retained. Again suppressing the $(\mathbf{x},\mathbf{p} ; t)$ notation, this can be seen in the equation of motion for $\mathbf{\tilde{B}}^{\text{el}}(\mathbf{x},\mathbf{p} ; t)$, 
\begin{align}
\frac{\partial \mathbf{\tilde{B}}^{\text{el}}}{\partial t} & = \{\!\!\{H^{\text{el}}_{\text{S}},A\boldsymbol{\tilde{\sigma}}^{\text{el}}_{\text{ss}}\}\!\!\}_{a} - \boldsymbol{\tilde{\sigma}}^{\text{el}}_{\text{ss}}\text{Tr}_{\text{el}}[\{\!\!\{H^{\text{el}}_{\text{S}},A\boldsymbol{\tilde{\sigma}}^{\text{el}}_{\text{ss}}\}\!\!\}_{a}] -\mathcal{L}^{\text{el}}\mathbf{\tilde{B}}^{\text{el}}. \label{eq: B TEV 4}
\end{align}
After solving this first-order differential equation and assuming that the system starts in the electronic steady state, such that $\mathbf{\tilde{B}}^{\text{el}}(\mathbf{x},\mathbf{p} ; 0) = 0$, time-dependent friction and diffusion tensors,
\begin{align}
\bar{\gamma}_{ij}(t - \tau) & = - \text{Tr}_{\text{el}}\left[\frac{\partial }{\partial x_{i}}e^{-\mathcal{L}^{\text{el}}(\mathbf{x})(t - \tau)}\frac{\partial \boldsymbol{\tilde{\sigma}}^{\text{el}}_{\text{ss}}(\mathbf{x})}{\partial x_{j}}\right] \\
\bar{D}_{ij}(t - \tau) & = \frac{1}{2}\text{Tr}_{\text{el}}\Big[\delta F_{i}e^{-\mathcal{L}^{\text{el}}(\mathbf{x})(t - \tau)}\left(\delta F_{j}\boldsymbol{\tilde{\sigma}}^{\text{el}}_{\text{ss}}  +\boldsymbol{\tilde{\sigma}}^{\text{el}}_{\text{ss}}\delta F_{j}\right)\Big],
\end{align}
are derived. They form part of the resulting non-Markovian Fokker-Planck equation,
\begin{widetext}
\begin{align}
 \frac{\partial A(t)}{\partial t} & = - \sum_{i}\frac{p_{i}}{m_{i}} \frac{\partial A(t)}{\partial x_{i}} + \sum_{i}F^{}_{i}(\mathbf{x}) \frac{\partial A(t)}{\partial p_{i}} + \sum_{ij}\int^{t}_{0} d\tau\bar{\gamma}_{ij}(t - \tau) \frac{\partial}{\partial p_{i}}\left(\frac{p_{j}}{m_{j}}A(\tau)\right) + \sum_{ij}\int^{t}_{0} d\tau \bar{D}_{ij}(t - \tau) \frac{\partial^{2}A(\tau)}{\partial p_{i}\partial p_{j}}, \label{eq: FP final app}
\end{align}
\end{widetext}
Note that this non-Markovian friction is not equivalent to that derived in Ref. \cite{Chen2019a}, where all non-Markovian effects are rigorously included and the full history of the classical trajectory is used.

\bibliography{Main_text_incl._fig.bib} 

\end{document}